\documentclass[reqno,12pt]{amsart}
\usepackage{graphics, color, amsmath, amsfonts, amssymb, amsthm, amscd}
\usepackage{cite}
\usepackage[latin1]{inputenc}

\usepackage{epsf}
\usepackage{graphicx}
\usepackage{psfrag}

\numberwithin{equation}{section}

\newtheorem{theorem}{Theorem}[section]

\newtheorem{lemma}{Lemma}[section]

\newtheorem{remark}{Remark}

\newtheorem{bigthm}{Theorem}   

\renewcommand{\eqref}[1]{(\ref{#1})}

\newcommand{\sumtwo}[2]{\sum_{\stackrel{#1}{ #2}}} 

\newcommand{\cA}{\ensuremath{\mathcal A}}
\newcommand{\cB}{\ensuremath{\mathcal B}}

\newcommand{\cD}{\ensuremath{\mathcal D}}
\newcommand{\cE}{\ensuremath{\mathcal E}}

\newcommand{\cG}{\ensuremath{\mathcal G}}
\newcommand{\cH}{\ensuremath{\mathcal H}}

\newcommand{\cL}{\ensuremath{\mathcal L}}

\newcommand{\cP}{\ensuremath{\mathcal P}}

\newcommand{\cR}{\ensuremath{\mathcal R}}

\newcommand{\cV}{\ensuremath{\mathcal V}}

\newcommand{\cZ}{\ensuremath{\mathcal Z}}

\newcommand{\frA}{\ensuremath{\mathfrak A}}

\newcommand{\frE}{\mathbf E}

\newcommand{\frG}{\ensuremath{\mathfrak G}}

\newcommand{\frI}{\ensuremath{\mathfrak I}}

\newcommand{\frR}{\ensuremath{\mathfrak R}}
\newcommand{\frS}{\ensuremath{\mathfrak S}}

\newcommand{\frg}{\ensuremath{\mathfrak g}}

\newcommand{\frm}{\ensuremath{\mathfrak m}}
\newcommand{\frn}{\ensuremath{\mathfrak n}}

\newcommand{\frt}{\ensuremath{\mathfrak t}}

\newcommand{\bbE}{{\ensuremath{\mathbb E}} }

\newcommand{\bbP}{{\ensuremath{\mathbb P}} }

\newcommand{\bbR}{{\ensuremath{\mathbb R}} }
\newcommand{\bbS}{{\ensuremath{\mathbb S}} }
\newcommand{\bbT}{{\ensuremath{\mathbb T}} }

\newcommand{\bbZ}{{\ensuremath{\mathbb Z}} }

\newcommand{\sfx}{{\sf x}}

\newcommand{\sfz}{{\sf z}}
\newcommand{\sfu}{{\sf u}}
\newcommand{\sfv}{{\sf v}}
\newcommand{\sfw}{{\sf w}}
\newcommand{\sft}{{\sf t}}

\newcommand{\sfe}{{\sf e}}

\newcommand{\sfp}{{\sf p}}
\newcommand{\sfq}{{\sf q}}
\newcommand{\sfr}{{\sf r}}

\newcommand{\sfO}{{\sf 0}}

\def\1{\ifmmode {1\hskip -5pt \rm{I}}
\else {\hbox {$1\hskip -3pt \rm{I}$}}\fi} 
\newcommand{\One}{1\hskip -3pt \rm{I}}

\newcommand{\df}{\stackrel{\Delta}{=}}

\newcommand{\wt}{\widetilde}
\newcommand{\Hsigmax}{\hat \Sigma^x}

\newcommand{\lb}{\left(}
\newcommand{\rb}{\right)}
\newcommand{\lbr}{\left\{}
\newcommand{\rbr}{\right\}}
\newcommand{\la}{\left\langle}
\newcommand{\rab}{\right\rangle}

\newcommand{\nup}{\nu^2}
\newcommand{\xip}{\xi^2}
\newcommand{\frmp}{{\frak m}^2 }

\newcommand{\lra}{\longleftrightarrow}
\def\nlra{\not \longleftrightarrow}
\newcommand{\slra}[1]{\stackrel{{#1}}{\lra}}
\newcommand{\nslra}[1]{\stackrel{{#1}}{\nlra}}
\newcommand{\ssim}[1]{\stackrel{{#1}}{\sim}}

\newcommand{\case}[1]{C{\small ASE}~#1\!\! }

\newcommand{\hsigmaz}{\hat\sigma^{\sfz}}
\newcommand{\hsigmax}{\hat\sigma^{\sfx}}
\newcommand{\Tr}{{\sf Tr}}

\newcommand{\dd}{{\rm d}}
\newcommand{\Cl}{\mathbf{\rm C}}

\newcommand{\Zdb}{\bbZ^\beta_\delta}

\newcommand{\bfJ}{{\mathbf J}}

\newcommand{\be}{\begin{equation}}
\newcommand{\ee}{\end{equation}}
\begin{document}

\title[Random Current Representation]{ Random Current Representation for Transverse
Field Ising Model}

\author{Nicholas Crawford }
\address{Department of Statistics\\
UC Berkeley; Berkeley, CA}
\email{}
\thanks{}
\author{
Dmitry Ioffe}
\address{
Faculty of Industrial Engineering\\
Technion, Haifa 3200, Israel}
\email{ieioffe\@@ie.technion.ac.il}
\thanks{This research was supported by a grant from G.I.F., 
the German Israeli Foundation for Scientific Research and Development}
\vskip 0.2in
\setcounter{page}{1}
\begin{abstract}
Random current representation (RCR) 
for transverse field Ising models (TFIM)  has been introduced in
\cite{ILN}.
This representation is a space-time version of the classical RCR 
exploited by Aizenman et. al.
\cite{Ai82, ABF, AF}. In this paper we formulate and prove corresponding space-time versions
 of  the classical switching lemma and show how they
generate various correlation inequalities. In particular we prove exponential decay of truncated
two-point functions at positive magnetic fields in $\sfz$-direction and address
the issue of the
sharpness of phase transition.
\end{abstract}
\maketitle

\section{The model and the results}
In what follows, we shall, for brevity, consider translation invariant models on $\mathbb Z^d$.
Specifically,
let $\bbT_N$ be the $d$-dimensional lattice torus
of linear size $N$ and $\bfJ = \lbr J_{ij}=J_{i-j}\rbr$
is  a finite range irreducible translation invariant interaction.
Let $h\geq 0$, $\rho >0$, $\lambda\geq 0$ and $0 \leq \beta \leq \infty$. The quantum Hamiltonian we are
going to consider is of the form,
\be
\label{eq:HamN}
-\cH_N = \frac{\rho}2\sum_{i,j}J_{ij}\hsigmaz_i\hsigmaz_j +h\sum_i\hsigmaz_i +
\lambda \sum_i
\Hsigmax_i .
\ee
Above $\Hsigmax = \lb I +\hsigmax\rb/2$, and $\hsigmaz$ and $\hsigmax$ are usual
 Pauli matrices,
\[
 \hsigmaz\, =\, \lb
\begin{array}{cc}
\ 1  &0\\
0 &-1
\end{array}
\rb .\quad\text{and}\quad
\hsigmax\, =\, \lb
\begin{array}{cc}
0  &1\\
1 &0
\end{array}
\rb
\]
Let us introduce the partition  function
\[
 \cZ_{\beta ,N} (h ,\rho ,\lambda ) \df {\rm e}^{- N^d\beta \lb
\rho 
{\bar{J}} +h +\lambda\rb }
\Tr\lb {\rm e}^{-\beta\cH_N}\rb
\]
where 
{$\bar{J}\df \sum_j
J_{i j}$} and we remark that this choice of normalization
is made so as to seamlessly introduce certain stochastic integral representations below.
Mean values of various local observables are denoted as $\langle\cdot\rangle_{\beta ,N}$.
For instance,
\begin{equation*}
\begin{split}
&\langle \hsigmaz_i\rangle_{\beta ,N} =
\frac{\Tr\lb \hsigmaz_i{\rm e}^{-\beta\cH_N}\rb }{\Tr\lb {\rm e}^{-\beta\cH_N}\rb}, \quad
\langle \Hsigmax_i\rangle_{\beta ,N} =
\frac{\Tr\lb \Hsigmax_i{\rm e}^{-\beta\cH_N}\rb }{\Tr\lb {\rm e}^{-\beta\cH_N}\rb},
\\
&\text{or, for $i\neq j$, }\quad
\langle\hsigmaz_j \Hsigmax_i\rangle_{\beta ,N} =
\frac{\Tr\lb\hsigmaz_j \Hsigmax_i{\rm e}^{-\beta\cH_N}\rb }{\Tr\lb {\rm e}^{-\beta\cH_N}\rb} .
\end{split}
\end{equation*}
Most of the results which we shall derive in the sequel hold uniformly in $\beta <\infty$
and/or in  $N$.
Whenever this {is the} the case we shall omit the corresponding sub-index.
Note that in many cases  uniformity in $\beta <\infty$ implies extensions
of the corresponding properties to the
ground state $\beta =\infty$.
\smallskip

Important quantities to be considered here are the $\sfz$-magnetization:
\[
 M_{\beta ,N} \lb h ,\rho ,\lambda \rb  = \langle \hsigmaz_i\rangle_{\beta ,N},
\]
and the truncated two-point functions,
\[
 \langle \hsigmaz_i ;\hsigmaz_j \rangle_{\beta ,N}\df \langle \hsigmaz_i
\hsigmaz_j \rangle_{\beta ,N} -
\langle \hsigmaz_i \rangle_{\beta ,N}\langle  \hsigmaz_j \rangle_{\beta ,N}, \ \
 \langle \hsigmaz_i ;\Hsigmax_j \rangle_{\beta ,N}\ \ \text{and}\ \
 \langle \Hsigmax_i ;\Hsigmax_j \rangle_{\beta ,N}.
\]
Our two main results are:
\begin{bigthm}
\label{Thm:Decay}
 For every 
$h >0$, {$\lambda\geq 0$ and $\rho\geq 0$} there exists $c_1 = c_1 (h, {\lambda ,
\rho } )>0$ and
$c_2 = c_2(h, {\lambda, \rho } ) <\infty$,  such that  
\be
\label{eq:ThmDecay}
\begin{split}
&0\leq \langle \hsigmaz_i ;\hsigmaz_j \rangle \leq c_2{\rm e}^{-c_1 |j-i |}, \quad
0\leq \langle \Hsigmax_i ;\Hsigmax_j \rangle \leq c_2{\rm e}^{-c_1 |j-i |}, \quad\\
&\text{and, for $i\neq j$,}\ \
- c_2{\rm e}^{-c_1 |j-i |}\leq \langle \hsigmaz_i ;\Hsigmax_j \rangle \leq 0.
\end{split}
\ee
By our convention the above results are claimed to be uniform in the
torus size $N$  and in $\beta <\infty$.
\end{bigthm}

\begin{bigthm}
 \label{Thm:DifIn}
Uniformly in
$h>0 $, $\rho >0$ and $\lambda >0$ the following differential inequalities hold:
\be
\label{eq:ThmDifIn1}
 M (h ,\rho ,\lambda )\leq h\frac{\partial M}{\partial h} + M^3 +
 M^{2}
\rho \frac{\partial M}{\partial\rho} - 2 \lambda M^2 \frac{\partial M}{\partial\lambda},
\ee
and,
\be
\label{eq:ThmDifIn2}
-\frac{\partial M}{\partial\lambda} \leq  \frac{M}{1-M^2}
\frac{\partial M}{\partial h}
\quad  \text{and}\ \
\frac{\partial M}{\partial\rho} \leq
{\bar{J}} M \frac{\partial M}{\partial h}
\ee
Again, by convention, the above inequalities are claimed to hold uniformly in $N$ and
 in $\beta <\infty$.
\end{bigthm}

In view of the  fundamental  techniques developed in \cite{AB, ABF},  differential inequalities
 \eqref{eq:ThmDifIn1} and \eqref{eq:ThmDifIn2} imply certain sharpness of phase transition
as the transverse field $\lambda$ and/or the inverse temperature $\beta$ are varied.  In particular, since $\hsigmaz$ and $\Hsigmax$ do not commute, the uniformity of our estimates in $\beta$ imply that taking $\beta \rightarrow \infty$, these inequalities still hold and can be used to derive a genuine quantum phase transition, albeit the fact that we derive it using
a somewhat  classical re-interpretation of the model (see Section 5).   In principle, since the model in question
could be considered as the strong coupling limit of $(d+1)$-dimensional classical Ising
models \cite{CKP, AKN}, Theorem~\ref{Thm:DifIn} could be attempted as a limiting
conclusion from the result of \cite{ABF}. 

The point of this paper, however, is 
to try to understand something new; that is to
develop general and robust  stochastic
geometric description of  quantum systems, hopefully also yielding simpler, or at least
alternative, proofs even in the  classical case of $\lambda =0$.
 In particular,
the conclusions of both the theorems above will become rather transparent in the
stochastic geometric context which we develop here.

The rest of this paper is organized as follows.  Section \ref{S:Geom} introduces a recasting of the transverse Ising model in a useful probabilistic language.  Further, we set down various geometric notions for this recasting which form the basis of our proofs of Theorem~\ref{Thm:Decay} and \ref{Thm:DifIn}.  Section \ref{S:Switch} applies these notions to the truncated correlation functions appearing in Theorem~{\ref{Thm:Decay}.  The resulting expressions may be seen as generalizing the results of the classical \textit{Switching Lemma} employed in \cite{Ai82, ABF, AF}. 
Section~\ref{S:DiffIn} provides a derivation of Theorem~\ref{Thm:DifIn}.
Section~\ref{S:EXPDEC} analyzes  expressions 
for truncated correlations  to obtain a proof of Theorem \ref{Thm:Decay}. 
 Finally, at the end of Section~\ref{S:EXPDEC} we briefly address the implications
 for {a} quantum phase transition in the ground state $\beta =\infty$.
\smallskip

\noindent

\smallskip
{\bf A Bibliographical Remark.} 
Shortly after the first draft  of this work was posted on the web,
there appeared \cite{BG}.  
The authors of \cite{BG} draw motivation from 
a parity calculus via strong coupling limits for classical RCR, and they 
develop what they call ``random-parity representation'' for TFIM. 
 The paper \cite{BG} 
%
contains very similar formulations and 
proofs of  the corresponding switching lemma and of the differential inequalities.
The following bibliographical remark is due:

(a) Although it might look ostensibly different, the  random-parity  representation of
\cite{BG} can be readily derived (see Remark~\ref{rem:RP} below) from the RCR 
which was introduced in \cite{ILN} and which we use here.
\cite{ILN} is a transcript of 
lectures given
at Prague's Probability school in 2006.

(b)  A simple  example of  the application  to TFIM 
 of the classical switching lemma  via limiting parity calculus 
appears in the  Appendix of \cite{CCIL}. This computation was described  to
one of the authors of \cite{BG} in the summer of 2007.  
There is a long way from this computation to the full ``quantum'' switching 
 lemma for TFIM, and we had no 
communications with the authors of \cite{BG} regarding the matter since then.  In fact,  
up to the last moment we did not know that the authors of \cite{BG} were working on 
 random-current type representations for TFIM.

\section{Stochastic geometry of the model}
\label{S:Geom}
The stochastic geometric approach to quantum models via {the} Lie-Trotter product
expansion  in the imaginary
time variable (additional dimension) and a subsequent
classical re-interpretation {was} introduced in \cite{Gi69}.   An important
mile-stone along these lines is the seminal paper \cite{AN94}.
{The approach expounded upon in that paper} has many degrees
of freedom in the sense that one can experiment with numerous decompositions of the Hamiltonian and with the basis in
which the Lie-Trotter expansion is performed to achieve different representations.  

We shall skip the derivation of the representation of interest  {in the present context} and proceed
directly to its probabilistic description.  We refer the interested reader to \cite{ILN} where
{the {\em quantum random current representation} we are using here was introduced and 
where}  various other stochastic
geometric descriptions {of}  the transverse field Ising model are discussed at length. 
\smallskip

To each site $i\in\bbT_N$ one attaches a copy $\bbS_\beta^i$ of the circle $\bbS_\beta$ of
circumference $\beta$. In the ground state case $\beta = \infty$, $\bbS_\infty\df \bbR$.
The resulting $(d+1)$-dimensional state space of the model is $ \frS_N\cup\frg$, where,
\[
 \frS_N\df  \cup_{i\in\bbT_N}\bbS_\beta^i,
\]
and $\frg$ is an artificial ``ghost site''.
The parameters $h, \bfJ$ and $\lambda$ enter the picture in the following fashion:  Consider
graphs  $\cG_N = \lb \cV_N ,\cE_N\rb$ with the vertex set $\cV_N = \bbT_N \cup\frg$,
and edge {set} $\cE_N = \cE_N^0 \cup\cE_N^\frg$ which comprise either
edges $\sfe = (i,j )\in\cE_N^0$ with $i,j\in\bbT_N$ and $J_{i-j} >0$, or 
$\sfe = (i,\frg )\in\cE_N^\frg$ with $i\in\bbT_N$.
As {above}, we omit the sub-index $N$ whenever it has no impact on the corresponding definition
or claim.
Let us define the following families  of {\em independent} Poisson
point processes on $\bbS_\beta$:
\subsubsection{Processes of flips} With each $\sfe\in\cE_N$ we associate a Poisson process
$\xi_\sfe$ which has intensity ${\rho}J_{i-j}$ if $\sfe = (i,j )$ and intensity
$h$ if $\sfe = (i ,\frg )$.
\subsubsection{Processes of marks} With each $i\in\bbT_N$ we associate a Poisson
process $\frm_i$ of intensity ${\lambda}$

In the sequel we shall denote the corresponding product measure as $\bbP \lb \dd \xi , \dd\frm\rb$.
In particular, for notational convenience, whenever there is no confusion the dependence on $(\beta, {\mathbf J}, h, \rho, \lambda)$ will be suppressed.
To write down the random current representation we still need to introduce the notion
of labels:
\subsubsection{Labels} Labels $\nu$ are piece-wise constant maps
$\nu :\frS_N\mapsto \lbr r,l\rbr$. Here $r$ and $l$ are just two symbols, which, if one traces
the
original derivation of \cite{ILN},
are related to the one particle eigenfunctions in the
transverse $\sfx$-basis.
  Given a realization $\lb \xi ,\frm \rb$ of the Poisson point processes
and a finite subset  $A\subset\frS$, let us say that a label $\nu$  is compatible 
(see Figure~1
)-- which will be denoted by
$\nu\ssim{A}\lb \xi ,\frm\rb$ --  if
\vskip 0.1cm

(1) $\nu_i$ has a jump at $\sfu$ for every $\sfu\in A$.

(2) All other jumps of $\nu$ happen at arrival times of $\xi$: For
$\sfe = (i ,\frg )$, an arrival of $\xi_{\sfe}$
 {\em enforces}
 a flip of $\nu_i$, and, similarly,  an arrival of $\xi_{ij}$ {\em enforces} a
simultaneous flip of
$\nu_i$ and $\nu_j$.

(3) For each $i$, $\nu_i (t) = r$ at each arrival time $t$  of $\frm_i$

To facilitate the notation we shall drop $A$ from $\nu\ssim{A}\lb \xi ,\frm\rb$ whenever
$A=\emptyset$.

\begin{figure}[h] 
\label{fig:RCR} 
\begin{center} 
\psfragscanon 
\psfrag{L}[l]{$\Lambda$}
\psfrag{b}{$\beta$} 
\psfrag{g}{$\frg$}
\psfrag{t=0}{$0$}
\psfrag{t}{$t$}
\psfrag{A}{Arrivals of flips $\xi_{i,j}$}
\psfrag{B}{Arrivals of flips $\xi_{i\frg}$}
\psfrag{C}{Arrivals of marks  $\frm_{i}$}
\psfrag{r}{$r$}
\psfrag{l}{$\ell$}
\psfrag{1}{$1$}
\psfrag{2}{$2$}
\psfrag{3}{$3$}
\psfrag{4}{$4$}
\psfrag{5}{$5$}
\psfrag{6}{$6$}
\includegraphics[width=6in]{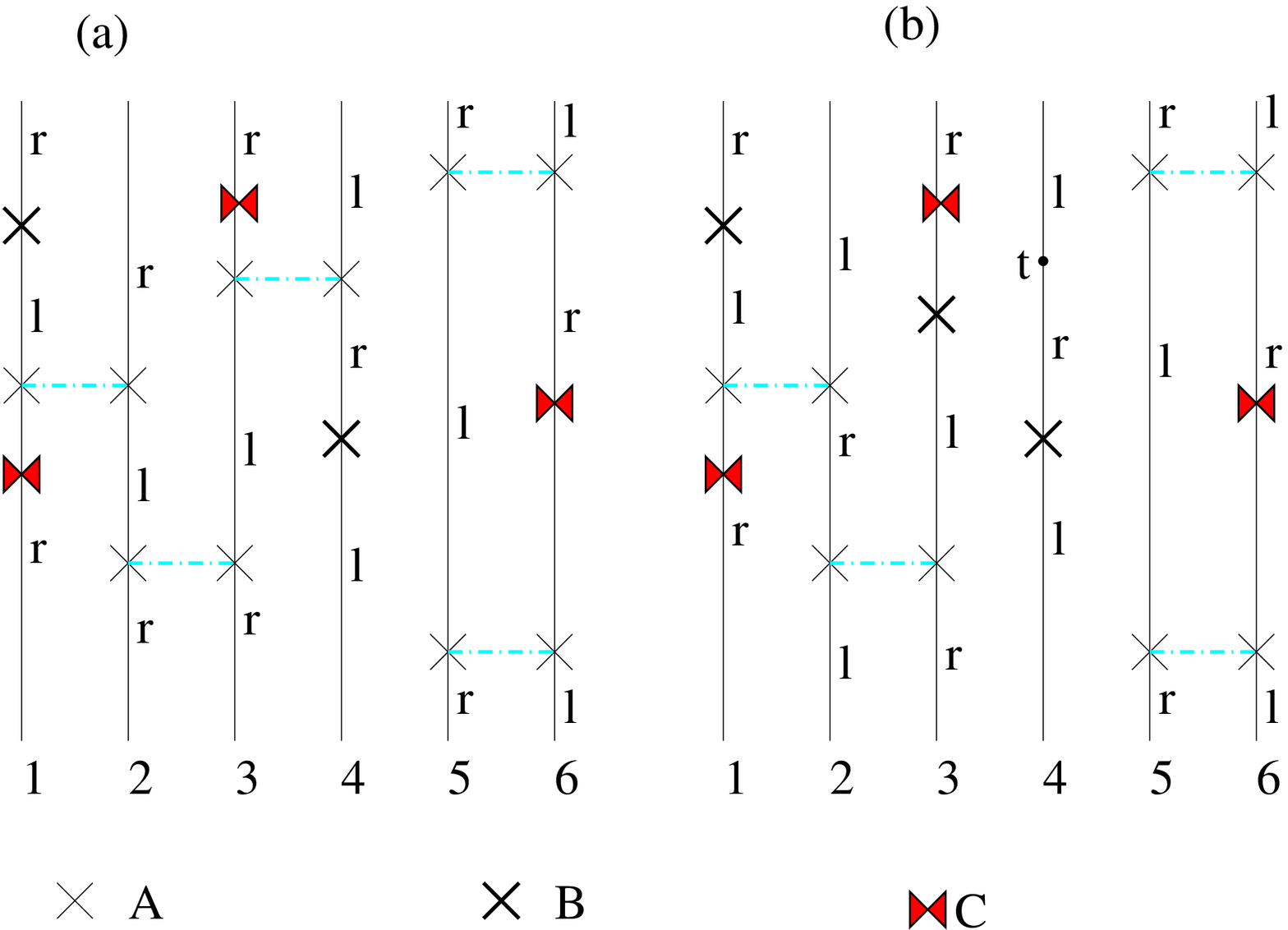}
\end{center} 
\caption{Poisson processes of arrivals and compatible labels on 
\newline $\frS = \cup_1^6\bbS_\beta^i$:\ \ 
\quad (a) $\nu\sim (\xi, \frm)$\quad (b)  $\nu\ssim{(4,t)}(\xi ,\frm )$ } 
\end{figure}

 \subsubsection{Representation Formulas}
The following formulas are established in \cite{ILN}:
For the partition function (and $\beta <\infty$),
\begin{equation}
\label{pf}
 \cZ_N =  \int
\bbP \lb \dd\xi ,\dd\frm\rb
\sum_{\nu\sim (\xi ,\frm )}{\bf 1} .
\end{equation}
\begin{remark}
 \label{rem:RP}
Integrating out the process of marks $\frm$ and calling $r$ ``even'' and 
$l$ ``odd'',  one recovers the ``random parity'' representation of 
\cite{BG}.
\end{remark}

Given $\sfu = (i,t )$ define
\[
\hsigmaz_\sfu = {\rm e}^{-t\cH}\hsigmaz_i {\rm e}^{t\cH} \ \ \text{and, accordingly,}\ \
\Hsigmax_\sfu = {\rm e}^{-t\cH}\hsigmax_i {\rm e}^{t\cH}.
\]
Note here that the signs match the imaginary time rotation of the quantum
 evolution.
{For one- and two-point functions in the $\sfz$ component of spin}:
%
%
\begin{equation}
\label{1ptf}
 \la\hsigmaz_\sfu\rab  = \frac1{\cZ }
\int
\bbP\lb \dd\xi ,\dd\frm\rb
\sum_{\nu\ssim{\sfu} (\xi ,\frm )}{\bf 1} .
\end{equation}
For the two-point function,
\begin{equation}
\label{2ptf}
 \la\hsigmaz_\sfu\hsigmaz_\sfv\rab  = \frac1{\cZ }
\int
\bbP\lb \dd\xi ,\dd\frm\rb
\sum_{\nu\ssim{\sfu, \sfv} (\xi ,\frm )}{\bf 1} .
\end{equation}
In fact, it is straightforward to check that similar formulas hold for $\sfx$-observables
and mixed two-point functions (see \cite{ILN} for details):  Namely,
\begin{equation}
\label{ptfx}
\begin{split}
 &\la\Hsigmax_\sfu\rab  = \frac1{\cZ}
\int
\bbP\lb \dd\xi ,\dd\frm\rb
\sum_{\nu\sim (\xi ,\frm )}\One_{\lbr \nu (u ) = r\rbr} ,\\
&
\la\Hsigmax_\sfu \Hsigmax_\sfv\rab  = \frac1{\cZ}
\int
\bbP\lb \dd\xi ,\dd\frm\rb
\sum_{\nu\sim (\xi ,\frm )}\One_{\lbr \nu (\sfu ) = r\rbr}\One_{\lbr \nu (\sfv ) = r\rbr},
\end{split}
\end{equation}
and,  for $\sfu\neq \sfv$,
\begin{equation}
\label{ptfzx}
\la\hsigmaz_\sfu \Hsigmax_\sfv\rab  = \frac1{\cZ}
\int
\bbP\lb \dd\xi ,\dd\frm\rb
\sum_{\nu\ssim{\sfu} (\xi ,\frm )} \One_{\lbr \nu (\sfv ) = r\rbr} .
\end{equation}
{Note that once these formulas are available with $\sfu \neq \sfv$, they may be extended by continuity to the appropriate limiting correlation functions.  We do not state them here as they will not appear in our derivations below.}

\subsubsection{Intervals, paths and replicas} Let $(\xi, \frm )$ be a realization
of the Poisson processes {introduced in the previous section}, $A$ a finite subset of 
$\frS$ and let $\nu$ be a
compatible label $\nu\ssim{A} \lb \xi ,\frm\rb$. An interval of $\nu$ is
a maximal connected component  $I= \lb \sfu ,\sfv\rb$ of some  $\bbS_\beta^i$ on which $\nu_i$ is
constant. A path $\cP$ of $( \nu , \xi , \frm)$ is an ordered  sequence
$\frI_1, \frI_2, \dots, \frI_n$, where $\frI_l$
is either an interval or a ghost site $\frg$ and,
\begin{enumerate}
 \item If $\frI_l = \lb \sfu_l ,\sfv_l\rb $ and $\frI_{l+1} = \lb \sfu_{l+1} ,\sfv_{l+1}\rb$ then either
$\sfv_l = \sfu_{l+1}$ or $\sfv_l = (i,t), \sfu_{l+1} = (j,t )$ and $t$ is an arrival time of
$\xi_{ij}$.
\item If $\frI_l = \lb \sfu_l ,\sfv_l\rb $, $\sfv_l = (j,t )$ and $\frI_{l+1}=\frg$, then $t$ is
an arrival time of $\xi_{i, \frg}$.
\item If $\frI_l =\frg$, $\frI_{l+1} = \lb \sfu_{l+1} ,\sfv_{l+1}\rb $ and
$\sfu_{l+1} = (j ,t )$, then $t$ is
an arrival time of $\xi_{j, \frg}$.
\item There could not be two successive ghost sites  $\frg$  in a path.
\end{enumerate}
A path $\cP = \lbr\frI_1,\dots ,\frI_n\rbr $ is said to be {\em ground} if it does not
contain $\frg$, {except possibly at the last step $\frI_n$.}
Finally, a path $\cP$ is said to be {\em left} if all the ground
intervals of $\cP$ bear $\nu$-label
$l$.

Let us define the set $\lbr \sfu \lra \sfv\rbr$ to be the collection of triples $(\xi, \frm, \nu)$ so that there exists a left
path with endpoints at $\sfu$ and $\sfv$ and the set $\lbr \sfu\slra{\sft}\sfv\rbr$ to be the collection of triples $(\xi, \frm, \nu)$ so that
exists a ground left path from $\sfu$ to $\sfv$. 
Note that ground left paths are self-avoiding and that
 there is a unique ground left path from $\sfu$ to $\frg$ whenever
$\nu\ssim{\sfu} (\xi ,\frm )$.  We shall denote this path by $\Cl^l(\sfu ,\frg )$ and
we shall use $\check{\Cl}^l  (\sfu ,\frg )$ for the union of
its ground intervals, that is for  $\Cl^l(\sfu ,\frg )\setminus\frg$.
\smallskip

Consider now two finite (and not necessarily disjoint) subsets $A,B\subset \frS$ and
two  copies $(\xi^1 ,\frm^1 ,\nu^1)$ and  $(\xi^2 ,\frm^2,\nu^2)$ such that
$\nu^1\ssim{A} (\xi^1 ,\frm^1 )$ and $\nu^2\ssim{B} (\xi^2 ,\frm^2 )$. We shall denote the
combined processes of flips and marks as
$( \eta ,\frn )\df (\xi^1\cup \xi^2 ,\frm^1 \cup\frm^2 )$,
where the union is understood in the
coordinate wise sense, e.g. $\eta_{ij} = \xi^{1}_{ij}\cup\xi^{2}_{ij}$. In all considerations
below the processes $(\xi^1 ,\frm^1 )$ and $(\xi^2 ,\frm^2 )$ are independent. Consequently,
$( \eta ,\frn )$ is just a collection of independent Poisson processes of arrivals with
double intensities. Furthermore, given a realization $(\eta ,\frn )$,
the conditional distribution of $(\xi^1 ,\frm^1 )\subseteq (\eta ,\frn )$ is uniform {with point mass} 
\be
\label{eq:CondD}
 \lb \frac12\rb^{\# (\eta ) +\# (\frn )}\df
\lb \frac12\rb^{\sum_{\sfe}\eta_{\sfe} [\bbS_\beta ]
+\sum_i \frn_i [ \bbS_\beta ]} .
\ee
Note that given $\eta$ and the locations of the discontinuities of $(\nu^1, \nu^2)$, the 
{ arrivals  of} $(\xi^1, \xi^2)$ may be recovered.  However it is not usually possible to reconstruct  $(\frm^1, \frm^2)$ from $\frn$ even knowing the values of $(\nu^1, \nu^2)$.

{Let us introduce geometric notions for \textit{pairs} of configurations, extending our previous definitions}. 
 It will be convenient to make definitions relative to a fixed finite 
subset  $G\subset\frS$. 
An interval $\frI$ of $(\nu^1 ,\nu^2 )$ is
a maximal connected component  $I= \lb \sfu ,\sfv\rb$ of some  $\bbS_\beta^i$, on which
 {\em both} labels  $\nu^1$ and $\nu^2$ are
constant and which does not contain points from $G$. 
A path $\cP$ of $( \nu^1, \nu^2, \eta , \frn)$ is an ordered  sequence
$\frI_1, \frI_2, \dots, \frI_n$, where $\frI_l$
is either an interval or a ghost site $\frg$ and,
\begin{enumerate}
 \item 
If $\frI_l = \lb \sfu_l ,\sfv_l\rb $ and $\frI_{l+1} = \lb \sfu_{l+1} ,\sfv_{l+1}\rb$ then either
$\sfv_l = \sfu_{l+1} = (i ,t)$, and then either $(i ,t)\in G$ or $t$ is an arrival time 
of $\eta_{i,\frg }$;  or, otherwise, 
$\sfv_l = (i,t), \sfu_{l+1} = (j,t )$ and $t$ is an arrival time of
$\eta_{ij}$.
\item If $\frI_l = \lb \sfu_l ,\sfv_l\rb $, $\sfv_l = (j,t )$ and $\frI_{l+1}=\frg$, then $t$ is
an arrival time of $\eta_{j, \frg}$.
\item If $\frI_l =\frg$, $\frI_{l+1} = \lb \sfu_{l+1} ,\sfv_{l+1}\rb $ and
$\sfu_{l+1} = (j ,t )$, then $t$ is
an arrival time of $\eta_{j, \frg}$.
\item There can not be two successive ghost sites  $\frg$  in a path.
\item All {\em ground} intervals $\frI_l\subset \frS$ are disjoint.
\end{enumerate}
As before, a path $\cP = \lbr\frI_1,\dots ,\frI_n\rbr $ is said to be {\em ground} if it does not
contain $\frg$, with a possible exception of the last  step $\frI_n$.
A path $\cP = \lbr 
\frI_1, \frI_2, \dots, \frI_n\rbr$ is 
  said to be a {\em loop} if  either $\frI_1 = \frI_n =\frg$ or $\sfv_n = \sfu_1$. 
It is useful to keep in mind that the above  notions do not depend on the values of
{compatible labels} 
 $(\nu_1, \nu_2)$ or {arrivals of marks} $\frn$.  Rather, they only depend on the 
{arrivals of flips $\eta$}. 

On the other hand, we also consider an important notion which very much 
depends on the pair of configurations:
 Let us say that the interval $\frI$ is {\em blocked} if (see Figure~2) 
both $\nu^1$ and $\nu^2$ equal to $r$ on $\frI$ and, in addition, $\frn ( \frI ) >0$.
\begin{figure}[h] 
\begin{center} 
\psfragscanon 
\psfrag{L}[l]{$\Lambda$}
\psfrag{b}{$\beta$} 
\psfrag{g}{$\frg$}
\psfrag{t=0}{$0$}
\psfrag{t}{{\Large $t$}}
\psfrag{s}{{\Large $s$}}
\psfrag{A}{Arrivals of flips and marks $(\xi^1 ,\frm^1 )$ of the first replica}
\psfrag{B}{Arrivals of flips and marks $(\xi^2 ,\frm^2 )$ of the second replica}
\psfrag{C}{Arrivals of marks  $\frm_{i}$}
\psfrag{r}{$r$}
\psfrag{l}{$\ell$}
\psfrag{1}{$1$}
\psfrag{2}{$2$}
\psfrag{3}{$3$}
\psfrag{4}{$4$}
\psfrag{5}{$5$}
\psfrag{6}{$6$}
\includegraphics[width=4in]{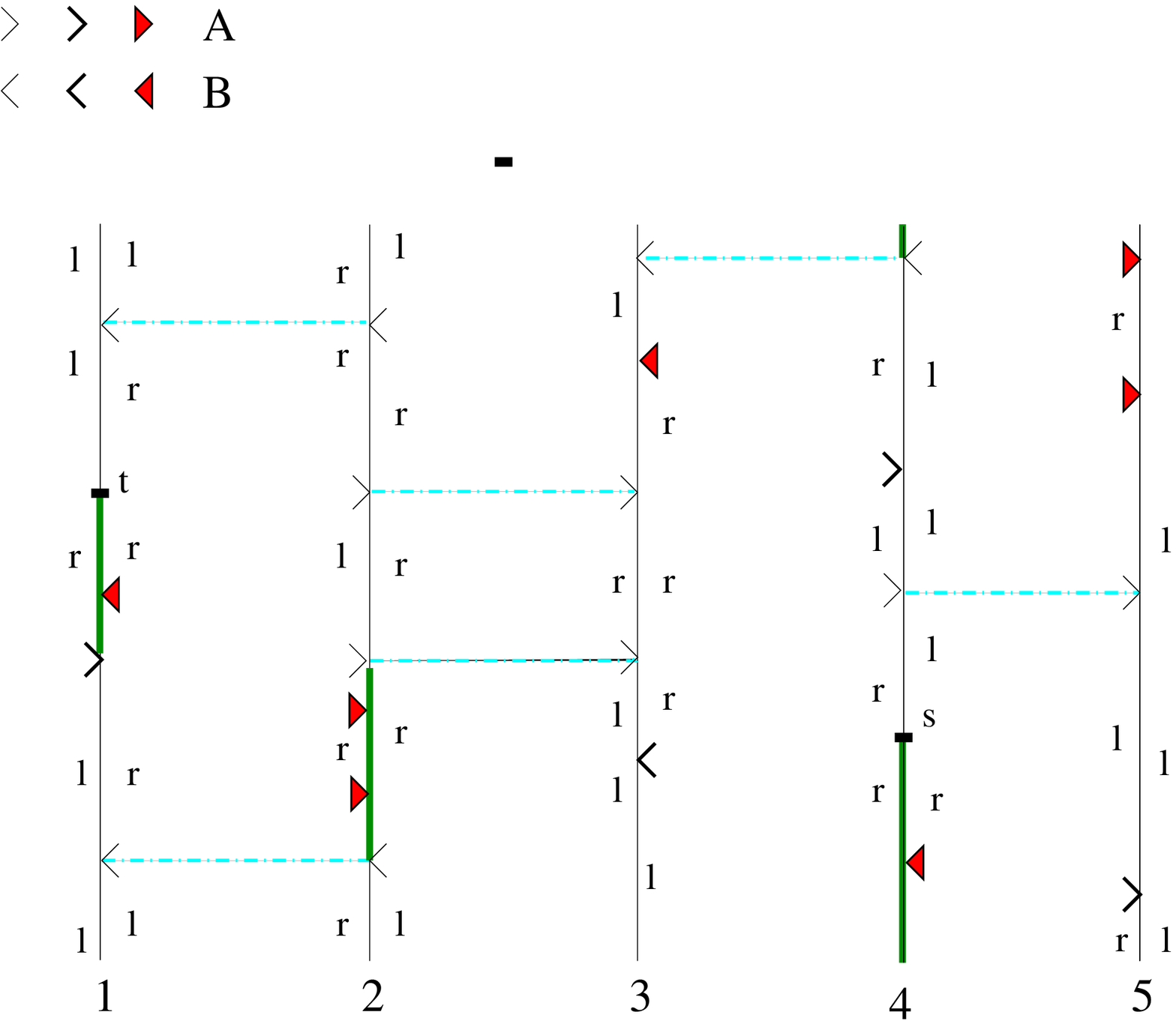}
\end{center} 
\caption{ Special set $G = \lbr (1,t ), (4,s )\rbr$: 
Blocked intervals for two replicas  $(\xi^1 ,\frm^1 )$, 
 $(\xi^2 ,\frm^2 )$ and two compatible labels $\nu^1\ssim{(1,t)} (\xi^1 ,\frm^1 )$, 
$\nu^2\ssim{(4,s)} (\xi^2 ,\frm^2 )$} 
\end{figure}

A path $\cP = \lbr\frI_1,\dots ,\frI_n\rbr $ is said to be {\em unblocked} if it does not
contain blocked intervals. We shall say that $\lbr \sfu\slra{*}\sfv\rbr$ if, for 
$G = \lbr \sfu ,\sfv\rbr$,  there
exists an unblocked path with end-points at $\sfu$ and $\sfv$, and we shall
write $\lbr \sfu\slra{*\frt}\sfv\rbr$ whenever there
exists
a ground unblocked path from $\sfu$ to $\sfv$.

\subsubsection{Basic Transformation} Let $\cP = \lb \frI_1 ,\dots, \frI_n\rb$ be an unblocked
 path of
$\lb\nu^1 ,\nu^2 ,\eta ,\frn\rb$ from $\sfu$ to $\sfv$. Obviously the labels $\nu^1$ and $\nu^2$ unambiguously
define the splitting $\eta = \xi^1\cup\xi^2$. Moreover, since $\cP$ is unblocked,
$\nu^1$ and $\nu^2$ unambiguously define the splitting of marks
 $\frn = \frm^1\cup\frm^2$ along $\cP$.

Make the following transformation of labels and marks on {\em each} of the
{\em ground}
 intervals $\frI$ of $\cP$:
\begin{enumerate}
 \item If the $(\nu^1 ,\nu^2)$ label of $\frI$ is $(l,r)$, then
flip it to $(r,l)$ and transfer all marks accordingly -- set $\frm^1(\frI) = \frm^2 (\frI )$ and set $\frm^2 (\frI) = 0$.
Perform the analogous procedure if the label is $(r,l)$.
\item If the label is $(l , l)$ then flip it to $(r,r)$. Accordingly, if the label is $(r, r)$, then
flip it to $(l, l)$. Note that in the latter case, since we are moving along an unblocked
path,  $\frn (\frI)$ has to be equal to zero, and no incompatibility arises.
\item Adjust $\xi^1$ and $\xi^2$ accordingly - those are, of course completely defined by the
labels (flips of the labels, to be precise).
\end{enumerate}
The above transformation, let us call it $\Phi_\cP$,  defines a map
\[
\lbr (\nu^1 ,\xi^1,  \frm^1 ), (\nu^2 ,\xi^2 ,\frm^2 )\rbr \mapsto
\lbr (\wt\nu^1 ,\wt\xi^1,  \wt\frm^1 ), (\wt\nu^2 ,\wt\xi^2, \wt\frm^2 )\rbr .
\]
The map $\Phi_\cP$ enjoys the following set of properties:
\begin{enumerate}
 \item It is invertible: Indeed just apply $\Phi_\cP$ once more to recover the original
data.
\item It does not change $\nu^1$ and $\nu^2$ labels and $\frm^1 ,\frm^2$ -marks
on intervals which do not belong to
$\cP$. In addition, the original and modified configurations
 have the same set of intervals
(defined by $\eta$, $\sfu$ and $\sfv$),
and $\Phi_\cP$ does not change the blocked/unblocked status of any of
those.
\item If $\nu^1\ssim{A} (\xi^1, \frm^1 )$ and $\nu^2\ssim{B} (\xi^2, \frm^2 )$, then
\be
\label{eq:LabelChange}
\wt\nu^1\ssim{A\Delta\lbr\sfu ,\sfv\rbr} (\wt\xi^1, \wt\frm^1 )\quad \text{and}\quad
\wt\nu^2\ssim{B\Delta\lbr\sfu ,\sfv\rbr} (\wt\xi^2, \wt\frm^2 ).
\ee
\item It is measure preserving: In view of \eqref{eq:CondD},  $(\xi^1 ,\frm^1)$
and $(\wt\xi^1 ,\wt\frm^1)$ have the same conditional weights.
\end{enumerate}
\subsubsection{Minimal paths} Most of the transformations we are going to perform will be
along {\em minimal} unblocked paths, {often satisfying additional geometric constraints}. Let us, therefore,  {define what we mean by \textit{minimal}.} First of all
given an unblocked path $\cP = \lb \frI_1, \dots ,\frI_n\rb$ define its length as
$|\cP |\df \sum_1^n |\frI_l |$, where $|\frI |$ is the Euclidean length if $\frI$ is a ground interval,
and, by definition, $|\frg |=0$. Consider now two replicas $\lb \xi^1 ,\frm^1\rb$, $\lb \xi^2 ,\frm^2\rb$
and a pair of compatible labels $\nu^1 ,\nu^2$. Let $\sfu ,\sfv\in \frS\cup \frg$ and assume that
there are unblocked paths from $\sfu$ to $\sfv$. Then the minimal path $\Cl^* (\sfu ,\sfv )$
satisfies,
\be
\label{eq:MinPath1}
|\Cl^* (\sfu ,\sfv )| \leq |\cP |\ \ \text{for any unblocked path  {$\cP$} from $\sfu$ to $\sfv$}.
\ee
It is easy to see that in general \eqref{eq:MinPath1} alone does not define $\Cl^* (\sfu ,\sfv )$
uniquely, and one needs to impose an additional rule in order to chose the minimal path
from a set of paths with the same {minimal} length. For example the following rule will do:
Write a coarse grained description of $\cP (\sfu ,\sfv ) = \frR_1 ,\dots ,\frR_m$, where
$\frR_l$ is either a ghost site $\frg$ or a maximal collection of successive ground intervals of
$\cP$ on some $\bbS_\beta^i$. Then for two unblocked paths $\cP = (\frR_1 ,\dots \frR_m)$ and
$\cP^\prime = (\frR_1^\prime  ,\dots, \frR_k^\prime) $ we shall say that $\cP\prec \cP^\prime$
if either $|\cP | < |\cP^\prime |$, or if the lengths are equal,
there exists $l$ such that
\be
\label{eq:MinPath2}
|\frR_i | = |\frR_i^\prime |\ \ \text{for $i=1,\dots,l-1$, but}\ \ |\frR_l | > |\frR^\prime_l |.
\ee
Then $\Cl^* (\sfu ,\sfv )$ is unambiguously defines as the unique unblocked path
from $\sfu$ to $\sfv$ which is $\prec$-less than any other unblocked path from
$\sfu$ to $\sfv$. {In other words, the minimal path, as we define it, 
 is the most conservative of all
the paths of the same minimal length: it tries to stay as much as possible on each 
subsequent spatial circle $\bbS_\beta$}

The important feature of the path transformation $\Phi$ which was introduced above is
(see Figure~3):
If $\Cl^* (\sfu ,\sfv )$ is the minimal path, then it remains so after
$\Phi_{\Cl^* (\sfu ,\sfv )}$
is performed. As a result, transformations along minimal paths are
well defined and invertible.

\begin{figure}[h] 
\begin{center} 
\psfragscanon 
\psfrag{L}[l]{$\Lambda$}
\psfrag{b}{$\beta$} 
\psfrag{g}{$\frg$}
\psfrag{t=0}{$0$}
\psfrag{t}{{\Large $t$}}
\psfrag{s}{{\Large $s$}}
\psfrag{A}{Arrivals of flips and marks $(\xi^1 ,\frm^1 )$ of the first replica}
\psfrag{B}{Arrivals of flips and marks $(\xi^2 ,\frm^2 )$ of the second replica}
\psfrag{C}{Arrivals of marks  $\frm_{i}$}
\psfrag{r}{$r$}
\psfrag{l}{$\ell$}
\psfrag{1}{$1$}
\psfrag{2}{$2$}
\psfrag{3}{$3$}
\psfrag{4}{$4$}
\psfrag{5}{$5$}
\psfrag{6}{$6$}
\includegraphics[width=6in]{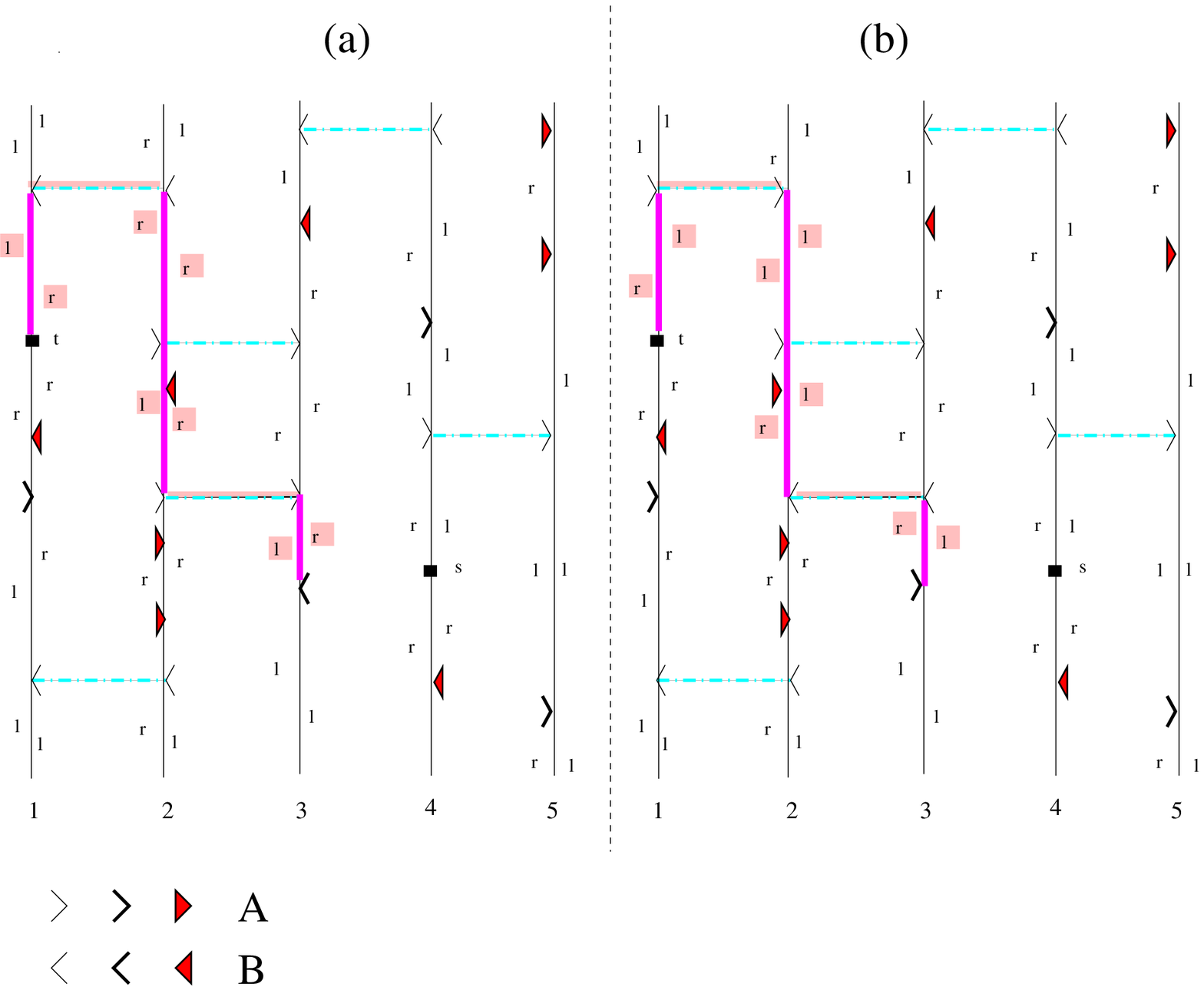}
\end{center} 
\caption{Two replicas  $(\xi^1 ,\frm^1 )$, 
 $(\xi^2 ,\frm^2 )$ and two compatible labels $\nu^1\ssim{\sfu } (\xi^1 ,\frm^1 )$, 
$\nu^2\ssim{\sfv } (\xi^2 ,\frm^2 )$, where $\sfu = (1,t)$ and $\sfv = (4,s )$.
\newline (a) Minimal unblocked path $\Cl^* (\sfu ,\sfv )$ from $\sfu $ to the ghost site $\frg$.
\newline (b) Basic transformation: New labels $\wt\nu^1\sim (\wt\xi^1 ,\wt\frm^1) $ and
$\wt\nu^2\ssim{\sfu ,\sfv } (\wt\xi^1 ,\wt\frm^1)$.  Labels  which are switched along
 the minimal path are shaded. Note that the flips and the marks  are switched accordingly.} 
\end{figure}

\section{Switching lemmas and related correlation inequalities}
\label{S:Switch}
Recall that  $\lb\xi^1 ,\frm^1\rb$ and $\lb\xi^2,\frm^2\rb$ are independent copies of
our Poisson processes of flips and marks, and that we use $\eta = \xi^1\cup \xi^2$,
$\frn = \frm^1\cup\frm^2$ for the combined processes.  {Let $\bbE$ denote the expectation with respect to two independent replicas
of Poisson processes of flips and marks; $\lb \xi^1, \frm^1\rb$ and $\lb\xi^2 ,\frm^2\rb$.}
In this section, we give exact formulae for the truncated correlations appearing in \eqref{eq:ThmDecay} and discuss the term
$\partial M/\partial\rho $ which appears in Theorem~\ref{Thm:DifIn}.
\subsubsection{Representation of $\la \hsigmaz_\sfu ; \hsigmaz_\sfv\rab$}
In view of \eqref{eq:CondD}
we can record \eqref{1ptf} in terms of two replicas as,
\begin{equation}
\label{prod1pf}
 \la\hsigmaz_\sfu\rab  \la\hsigmaz_\sfv\rab = \frac1{\cZ^2}
\int
\bbP \lb \dd\eta ,\dd\frn\rb
\lb \frac12\rb^{\# (\eta ) +\# (\frn )}\!\!\!\!
\sumtwo{\xi^1\cup\xi^2 =\eta}{\frm^1 \cup\frm^2 =\frn}
\sumtwo{\nu^1\ssim{\sfu} (\xi^1 ,\frm^2 )}{\nu^2\ssim{\sfv}(\xi^1 ,\frm^2 )}\!\!
{\bf 1} .
\end{equation}
Similarly, we can record \eqref{pf} and \eqref{2ptf} as,
\begin{equation}
\label{prod2pf}
\la\hsigmaz_\sfu \hsigmaz_\sfv\rab  =
\frac{\cZ\la\hsigmaz_i\hsigmaz_j\rab_\Lambda}{\cZ}
 = \frac1{\cZ^2}
\int
\bbP\lb \dd\eta ,\dd\frn\rb
\lb \frac12\rb^{\# (\eta ) +\# (\frn )}\!\!\!\!
\sumtwo{\xi^1\cup\xi^2 =\eta}{\frm^1 \cup\frm^2 =\frn}
\sumtwo{\nu^1\sim (\xi^1,\frm^1 )}{\nu^2\ssim{\sfu ,\sfv }(\xi^2 ,\frm^2 )}
\!\!
{\bf 1} .
\end{equation}
Let us have a closer look at \eqref{prod1pf}. The constraint $\nu^1\ssim{\sfu } (\xi^1 ,\frm^1 )$ implies that
there is a path
$\cP$ from $\sfu$ to $\frg$ such that $\nu^1\equiv l$ on $\cP$.
In particular this path $\cP$ must be
 unblocked. An analogous statement also applies with respect to $\sfv$ in the second replica.
Therefore, one can rewrite \eqref{prod1pf} as
\begin{equation}
\label{prod1pf_ub}
\begin{split}
 &\la\hsigmaz_\sfu\rab\la\hsigmaz_\sfv\rab \\
&= \frac1{\cZ^2}
\int
\bbP\lb \dd\eta ,\dd\frn\rb
\lb \frac12\rb^{\# (\eta ) +\# (\frn )}\\
&\qquad\times
\sumtwo{\xi^1\cup\xi^2 =\eta}{\frm^1 \cup\frm^2 =\frn}
\sumtwo{\nu^1\ssim{\sfu} (\xi^1 ,\frm^1 )}{\nu^2 \ssim{\sfv}(\xi^2 ,\frm^2 )}\!\!
\One_{\lbr \sfu\slra{*}\frg\rbr}\One_{\lbr \sfv\slra{*}\frg\rbr}.
\end{split}
\end{equation}
Similarly, one can rewrite \eqref{prod2pf} as,
\begin{equation}
\label{eq:prod2pf_ub}
\begin{split}
 &\la\hsigmaz_\sfu \hsigmaz_\sfv\rab \\
&= \frac1{\cZ^2}
\int
\bbP\lb \dd\eta ,\dd\frn\rb
\lb \frac12\rb^{\# (\eta ) +\# (\frn )}\!\!\!\!\!\times
\sumtwo{\xi^1\cup\xi^2 =\eta}{\frm^1 \cup\frm^2 =\frn}
\sumtwo{\nu^1\sim  (\xi^1 ,\frm^1 )}{\nu^2\ssim{\sfu ,\sfv}(\xi^2 ,\frm^2 )}\!\!
\One_{\lbr \sfu\slra{*}\sfv\rbr}.
\end{split}
\end{equation}
Let us fix a realization of $(\eta ,\frn )$. Define
$\cA_{\sfu ,\sfv}^\frg =\cA_{\sfu ,\sfv}^\frg (\eta ,\frn )$ to be the set of pairs
of objects $\lbr (\nu^1 , \xi^1 ,\frm^1  ),
(\nu^2 ,\xi^2 ,\frm^2  )\rbr$ which contribute to the double sum on the right hand side of
\eqref{prod1pf_ub}. Similarly let $\cA_{\sfu ,\sfv}$ be the set of pairs of objects (currents and labels) which contribute to the double sum on the right hand side of \eqref{eq:prod2pf_ub}. Each 
of the objects
in $\cA_{\sfu ,\sfv}^\frg$ contains an unblocked path, and hence the minimal unblocked path
 $\Cl^* (\sfu ,\frg )$ {from $\sfu$ to $\frg$}. We claim that the 
map , $\Phi \equiv \Phi_{\Cl^* (\sfu ,\frg )} : \cA_{\sfu ,\sfv}^\frg\mapsto \cA_{\sfu ,\sfv}$ is a measure preserving
injection.
This follows immediately from the properties of basic transformations and minimal paths.
 However, $\Phi$ is not onto: any couple of objects  in the image
$\lbr (\nu^1, \xi^1 ,\frm^1 ),
(\nu^2, \xi^2 ,\frm^2  )\rbr\in \Phi (\cA_{\sfu ,\sfv}^\frg)$ necessarily contains an unblocked path
from $\sfu$ to $\frg$. We have proved:
\begin{theorem}
\label{thm:Switchz}
 Truncated $\sfz$-correlation functions
satisfy the following version of the Switching Lemma:
\begin{equation}
\label{prod2pf_ub}
\begin{split}
 &\la\hsigmaz_\sfu ; \hsigmaz_\sfv \rab  \\
&= \frac1{\cZ^2}
\int
\bbP\lb \dd\eta ,\dd\frn\rb
\lb \frac12\rb^{\# (\eta ) +\# (\frn )}
\times
\sumtwo{\xi^1 \cup\xip =\eta}{\frm^1 \cup\frmp =\frn}
\sumtwo{\nu^1\sim  (\xi^1 ,\frm^1 )}{\nup\ssim{\sfu ,\sfv }(\xip ,\frmp )}\!\!
\One_{\lbr \sfu\nslra{*}\frg\rbr}\\
& =
\frac1{\cZ^2}\ \bbE \!\!\!\!
\sumtwo{\nu^1\sim  (\xi^1 ,\frm^1 )}{\nup\ssim{\sfu ,\sfv }(\xip ,\frmp )}\!\!
\One_{\lbr \sfu\nslra{*}\frg\rbr}.
\end{split}
\end{equation}
\end{theorem}

\subsubsection{Representation of $\la \Hsigmax_\sfu ; \Hsigmax_\sfv\rab$}
\label{S:x}
Consider two independent replicas $(\xi^1 , \frm^1 )$, $(\xi^2 , \frm^2 )$ and two
labels $\nu^1\sim (\xi^1 , \frm^1 )$ and $\nu^2\sim (\xi^2 , \frm^2 )$.
 Let us say that a couple of
labels $(\nu^1, \nu^2 )\in [(r,r)_{{\sfu}}, (r,l )_{{\sfv}}]$ if
$\nu^1 (\sfu )= r = \nu^1 (\sfv )$, $\nu^2 (\sfu )=r $ and
$\nu^2 (\sfv) =l$.  The events, $\lbr (\nu^1 , \nu^2 )\in [(r,l)_\sfu , (r,l )_\sfv]\rbr $,
$\lbr (\nu^1 , \nu^2 )\in [(l,l)_\sfu , (r,l )_\sfv]\rbr $ etc. (all together 16 events) are defined in a completely
similar fashion.
In terms of two replicas, the representation formulas
\eqref{ptfx} read as,
\begin{equation}
\label{eq:2repij1}
\begin{split}
\la\Hsigmax_\sfu \Hsigmax_\sfv\rab  = \frac1{\cZ^2}\, \bbE \!\!
\sumtwo{\nu^1\sim (\xi^1 ,\frm^1 )}{\nu^2\sim (\xi^2 ,\frm^2 )}
\Big( &\One_{\lbr (\nu^1 , \nu^2 )\in [(r,r)_\sfu , (r,r )_\sfv]\rbr}+
\One_{\lbr (\nu^1, \nup )\in [(r,r)_\sfu , (r,l )_\sfv ]\rbr}\\
& +  \One_{\lbr (\nu^1 , \nup )\in [(r,l)_\sfu , (r,r )_\sfv ]\rbr} +
\One_{\lbr (\nu^1 , \nup )\in [(r,l)_\sfu , (r,l )_\sfv ]\rbr }\Big).
\end{split}
\end{equation}
Similarly,
\begin{equation}
\label{eq:2repij2}
\begin{split}
\la\Hsigmax_\sfu\rab
\la \Hsigmax_\sfv\rab = \frac1{\cZ^2}\, \bbE \!\!\!\!
\sumtwo{\nu^1 \sim (\xi^1 ,\frm^1 )}{\nup\sim (\xip ,\frmp )}
\Big( &\One_{\lbr (\nu^1 , \nup )\in [(r,r)_\sfu , (r,r )_\sfv ]\rbr}+
\One_{\lbr (\nu^1 , \nup )\in [(r,l)_\sfu , (r,r )_\sfv ]\rbr}\\
& +  \One_{\lbr (\nu^1 , \nup )\in [(r,r)_\sfu , (l,r )_\sfv ]\rbr} +
\One_{\lbr (\nu^1 , \nup )\in [(r,l)_\sfu , (l,r  )_\sfv ]\rbr }\Big).
\end{split}
\end{equation}
Evidently,
\[
\bbE \!\!\!\!
\sumtwo{\nu^1\sim (\xi^1 ,\frm^1 )}{\nup\sim (\xip ,\frmp )}
\One_{\lbr (\nu^1 , \nup )\in [(r,r)_\sfu , (r,l )_\sfv ]\rbr} =
\bbE \!\!\!\!
\sumtwo{\nu^1\sim (\xi^1 ,\frm^1 )}{\nup\sim (\xip ,\frmp )}
\One_{\lbr (\nu^1 , \nup )\in [(r,r)_\sfu , (l,r )_\sfv ]\rbr} .
\]
Consequently, we arrive to the following representation for the truncated
two point function:
\begin{equation}
\label{eq:Rep2pf_x}
\la\Hsigmax_\sfu ; \Hsigmax_\sfv\rab  = \frac1{\cZ^2}\  \bbE \!\!\!\!
\sumtwo{\nu^1\sim (\xi^1 ,\frm^1 )}{\nup\sim (\xip ,\frmp )}
\lb  \One_{\lbr (\nu^1 , \nup )\in [(r,l)_\sfu , (r,l )_\sfv ]\rbr} -
\One_{\lbr (\nu^1 , \nup )\in [(r,l)_\sfu , (l,r )_\sfv ]\rbr} \rb .
\end{equation}

At this stage we proceed
 much along the lines of our proof of Theorem~\ref{thm:Switchz}.
Fix a realization
 of $\lb \eta ,\frn\rb$ and let $\cB_+ (\eta ,\frn )$ be the set of pairs of objects
$\lbr (\nu^1,  \xi^1 ,\frm^1  ) , (\nup ,\xip ,\frmp )\rbr$ which contribute to the sum
\[
\sumtwo{\xi^1\cup\xip =\eta}{\frm^1\cup\frmp = \frn}\
\sumtwo{\nu^1 \sim (\xi^1 ,\frm^1 )}{\nup\sim (\xip ,\frmp )}
\One_{\lbr (\nu^1 , \nup )\in [(r,l)_\sfu , (r,l )_\sfv ]\rbr}.
\]
Similarly, let
$\cB_- (\eta ,\frn )$ be the set of pairs of objects
$\lbr ( \nu^1 ,\xi^1 ,\frm^1 ) , (\nu^2, \xi^2 ,\frm^2 )\rbr$ which contribute to the sum
\[
\sumtwo{\xi^1\cup\xi^2 =\eta}{\frm^1\cup\frm^2 = \frn}\
\sumtwo{\nu^1\sim (\xi^1 ,\frm^1 )}{\nu^2\sim (\xi^2 ,\frm^2 )}
\One_{\lbr (\nu^1 , \nu^2 )\in [(r,l)_\sfu , (l,r )_\sfv ]\rbr} .
\]
An injective map $\Psi =\Psi_{\eta ,\frn}  :  \cB_- (\eta ,\frn )
\mapsto \cB_+ (\eta ,\frn )$ is
constructed as follows:  Any
\[
\lbr ( \nu^1, \xi^1 ,\frm^1  ) , (\nup, \xip ,\frmp)\rbr\in \cB_- (\eta ,\frn )
\]
contains an unblocked loop $\cL$
 from $\sfv$ to $\sfv$ such that
$\sfu\not\in\cL$. Indeed, such a loop may be constructed with $\nu^1\equiv l$. Now just choose
the minimal such loop (in the sense discussed above) and perform on this minimal loop the very same
surgery as in the Basic Transformation. Again, the property that the loop is
minimal is not changed under the surgery and hence $\Psi$ is invertible. On the other hand,
the image
set $\Psi \cB_- (\eta ,\frn )\subset \cB_+ (\eta ,\frn )$.

Geometrically, it is evident that $\cB_+\setminus\Psi\cB_-$ is characterized by the 
following condition:
A pair  $\lbr ( \nu^1 ,\xi^1 ,\frm^1 ) , (\nup, \xip ,\frmp  )\rbr$ from $\cB_+$ belongs
 to $ \cB_+\setminus\Psi\cB_-$
 if and only if any unblocked loop containing $\sfv$ also contains $\sfu$.  In this case, let
us say that $\sfu$ is {\em loop-pivotal} for $\sfv$.

We conclude:
\begin{theorem}
\label{thm:Switchx}
Truncated $\sfx$-correlation functions
satisfy the following version of the Switching Lemma:
\begin{equation}
\label{eq:Trunc2pt_x}
\la\Hsigmax_\sfu ; \Hsigmax_\sfv\rab  =
\frac1{\cZ^2}\
\bbE
\!\!\!\!
\sumtwo{\nu^1\sim (\xi^1 ,\frm^1 )}{\nup\sim (\xip ,\frmp )}
\One_{\lbr (\nu^1 , \nup )\in [(r,l)_\sfu , (r,l )_\sfv ]\rbr}\One_{\{\sfv \text{ is loop 
pivotal for $\sfu$}\}} .
\end{equation}
\end{theorem}

\subsubsection{Representation of cross-correlations}
As before, let $\bbE$ denote the expectation with respect to two independent replicas
of Poisson processes of flips and marks; $\lb \xi^1, \frm^1\rb$ and $\lb\xi^2 ,\frm^2\rb$.
With this notation we have (from \eqref{1ptf}, the first of \eqref{ptfx} and
\eqref{ptfzx}),
\be
\label{eq:Twoxz}
\la \hsigmaz_\sfu \Hsigmax_\sfv\rab = \frac1{\cZ^2}\
\bbE\!\!\!\!
\sumtwo{\nu^1\ssim{\sfu} (\xi^1 ,\frm^1 )}{\nu^2\sim (\xi^2 ,\frm^2 )}
\One_{\lbr\nu^1 (\sfv )= r\rbr} ,
\ee
and, accordingly, 
\be
\label{eq:prodxz}\la \hsigmaz_\sfu\rab
\la
\Hsigmax_\sfv\rab = \frac1{\cZ^2}\ \bbE
\sumtwo{\nu^1\ssim{\sfu} (\xi^1 ,\frm^1 )}{\nu^2\sim (\xi^2 ,\frm^2 )}
\One_{\lbr\nu^2 (\sfv)= r\rbr} .
\ee

Fix a realization
 of $\lb \eta ,\frn\rb$ and let $\cD_+ (\eta ,\frn )$ be the set of pairs of objects\newline
$\lbr (\nu^1,  \xi^1 ,\frm^1  ) , (\nup ,\xip ,\frmp )\rbr$ which contribute to the sum
\[
\sumtwo{\xi^1\cup\xip =\eta}{\frm^1\cup\frmp = \frn}\
\sumtwo{\nu^1 \ssim{\sfu} (\xi^1 ,\frm^1 )}{\nup\sim (\xip ,\frmp )}
\One_{\lbr \nup (\sfv ) = r\rbr} .
\]
Similarly, let
$\cD_- (\eta ,\frn )$ be the set of pairs of objects
$\lbr ( \nu^1 ,\xi^1 ,\frm^1 ) , (\xip ,\frmp , \nup )\rbr$ which contribute to the sum
\[
\sumtwo{\xi^1\cup\xip =\eta}{\frm^1\cup\frmp = \frn}\
\sumtwo{\nu^1\ssim{\sfu} (\xi^1 ,\frm^1 )}{\nup\sim (\xip ,\frmp )}
\One_{\lbr \nu^1 (\sfv ) =r\rbr} .
\]
Note now that any pair of objects 
$\lbr ( \nu^1 ,\xi^1 ,\frm^1 ) , (\nup, \xip ,\frmp  )\rbr \in \cD_-$
 contains an unblocked and hence the minimal unblocked path  $\Cl^{*, \not\, \sfv} (\sfu ,\frg )$
from
$\sfu$ to $\frg$ which {\em avoids} $\sfv$.
An injective map $\Omega =\Omega_{\eta ,\frn}  :  \cD_- (\eta ,\frn )
\mapsto \cD_+ (\eta ,\frn )$ is then
constructed as follows:
\begin{enumerate}
 \item  Perform the Basic Transformation along the minimal path $\Cl^{*,\not\, \sfv} (\sfu ,\frg )$.
\item Using the symmetry of replicas, rename the resulting
\[
\lb \wt\nu^1 ,\wt\xi^1 ,\wt\frm^1\rb \leftrightarrow
\lb \wt\nu^2,\wt\xi^2 ,\wt\frm^2\rb .
\]
\end{enumerate}
It is evident that $\cD_+\setminus\Omega\cD_-$ is characterized by the following condition:
A pair of objects  $\lbr ( \nu^1 ,\xi^1 ,\frm^1 ) , (\nup, \xip ,\frmp )\rbr$ from $\cD_+$ belongs
 to $ \cD_+\setminus\Omega\cD_-$ if and only if any unblocked path from $\sfu$ to the
ghost site $\frg$ contains $\sfv$.  Let us say that $\sfv$ is {\em pivotal} for $\sfu\slra{*}\frg$
if the latter condition holds. We have proved:
\begin{theorem}
\label{thm:Switchzx}
 Truncated cross-correlation functions
satisfy the following version of the Switching Lemma:
\begin{equation}
\label{eq:Trunc2pt_zx}
\la\hsigmaz_\sfu ; \Hsigmax_\sfv\rab =
 - \frac1{\cZ^2}\
\bbE\!\!\!\!
 \sumtwo{\nu^1\ssim{\sfu} (\xi^1 ,\frm^1 )}{\nup\sim (\xip ,\frmp )}
\One_{\lbr \nu^2 (\sfv ) = r\rbr}
\One_{\lbr \text{ $\sfv$ is pivotal for $\sfu\slra{*}\frg$ }\rbr}
\end{equation}
\end{theorem}
Note that the following (straightforward) generalization of Theorem~\ref{thm:Switchzx}
holds.  Let $G= \lbr \sfv_1, \dots, \sfv_l ,\sfv_l, \dots ,\sfv_{l+k}\rbr$ be a
finite subset of $\frS$ which is {\em time-ordered} in the following sense: The coordinates
 $\sfv_q = (\i_q , t_q )$ satisfy $t_q < t_p$ whenever $q<p$.  Let $\sfu = (i, t)$  be such
that $t_l < t <t_{l+1}$.  Then, the truncated cross-correlation
$\la \hsigmaz_\sfu ;\prod_1^{l+k} \Hsigmax_{\sfv_q}\rab$ is defined as
\[
 \la \hsigmaz_\sfu ;\prod_1^{l+k} \Hsigmax_{\sfv_q}\rab
=
 \la\prod_1^{l} \Hsigmax_{\sfv_q} \hsigmaz_\sfu \prod_{l+1}^{l+k} \Hsigmax_{\sfv_q}\rab
-
 \la \hsigmaz_\sfu \rab
 \la \prod_1^{l+k} \Hsigmax_{\sfv_q}\rab .
\]
We have,
\be
\label{eq:Crossmany}
\la \hsigmaz_\sfu ;\prod_1^{l+k} \Hsigmax_{\sfv_q}\rab =
- \frac1{\cZ^2}\
\bbE\!\!\!\!
 \sumtwo{\nu^1\ssim{\sfu} (\xi^1 ,\frm^1 )}{\nup\sim (\xip ,\frmp )}
\prod_1^{l+k}
\One_{\lbr \nu^2 (\sfv_q ) = r\rbr}
\One_{\lbr \text{ $G$ is pivotal for $\sfu\slra{*}\frg$ }\rbr} .
\ee
\subsubsection{Further correlation inequalities}
{In the classical
case (see e.g. \cite{Ai82, AF, ABF, Shl})  random current representations of correlations
generate a variety of correlation inequalities. In fact, the morphology in the quantum
case is even richer and this issue will be systematically addressed in a future paper
 \cite{CIF}.}  Here we shall focus only on such inequalities which are needed for
proving our main results.

{Partial derivatives with respect to the parameters $(h, \lambda, \rho)$ of the magnetization 
$M = \langle \hsigmaz_0 \rangle$ are related to truncated correlations  in the following
way: Fix the origin
$0$ of $\bbT_N$ and let $\sf0\in\frS$ be the point with the space time
coordinates $\sfO = (0,0)$. In view of (space and time) translation invariance} it is of course
inessential how we fix $\sfO$. Then,
\be
\label{eq:derivatives}
\begin{split}
&\frac{\partial M}{\partial h} = \sum_{i\in\bbT_N}\int_0^\beta
\la\hsigmaz_\sfO ; \hsigmaz_{(i,t)}\rab\dd t ,\quad
\frac{\partial M}{\partial \rho} =
\sum_{(i,j): J_{i-j}>0} 
\frac{J_{i j}}{2}\int_0^\beta
\la\hsigmaz_\sfO ; \hsigmaz_{(i,t) }\hsigmaz_{(j,t) }\rab\dd t,\\
&\text{and}\quad
\frac{\partial M}{\partial \lambda} = \sum_{i\in\bbT_N}\int_0^\beta
\la\hsigmaz_\sfO ; \Hsigmax_{(i,t)}\rab\dd t .
\end{split}
\ee

Random current representations for the $\sfz$ and cross-correlations were already given above.
Let us therefore turn to $\la\hsigmaz_\sfO ; \hsigmaz_{(i,t) }\hsigmaz_{(j,t) }\rab$ terms.
In order to facilitate the notation set $\sfw = (i,t)$ and $\sfz = (j,t)$.  The random current
 representation of
\[
 \la \hsigmaz_\sfO  \hsigmaz_{\sfw }\hsigmaz_{\sfz }\rab =
 \la \hsigmaz_\sfO   \hsigmaz_{\sfw}\hsigmaz_{\sfz }\rab\frac{\cZ}{\cZ} =
\frac{1}{\cZ^2}\bbE\sumtwo{\nu^1\ssim{\lbr\sfO ,\sfw ,\sfz\rbr}\lb\xi^1 ,\frm^1\rb}
{\nu^2\sim \lb\xi^2 ,\frm^2\rb} \1,
\]
is straightforward. Consider now,
\[
 \la\hsigmaz_{\sfw }\hsigmaz_{\sfz }\rab\la\hsigmaz_\sfO\rab
=
\frac{1}{\cZ^2}\bbE\sumtwo{\nu^1\ssim{\lbr\sfw ,\sfz\rbr}\lb\xi^1 ,\frm^1\rb}
{\nu^2\ssim{\sfO} \lb\xi^2 ,\frm^2\rb} \1 .
\]
Each pair of triples $\lbr \lb \nu^1, \xi^1 ,\frm^1\rb,\lb \nu^2, \xi^2 ,\frm^2\rb \rbr$
 which contributes to the latter integral contains an unblocked path from $\sfO$ to $\frg$.
Performing our Basic Transformation along the minimal such path, we infer,
 \[
 \la\hsigmaz_{\sfw }\hsigmaz_{\sfz }\rab\la\hsigmaz_\sfO\rab = 
\frac{1}{\cZ^2}\bbE\sumtwo{\nu^1\ssim{\lbr\sfO ,\sfw ,\sfz\rbr}\lb\xi^1 ,\frm^1\rb}
{\nu^2\sim  \lb\xi^2 ,\frm^2\rb} \One_{\lbr \sfO\slra{*}\frg\rbr} .
\]
Consequently,
\be
\label{eq:TripleCor}
 \la \hsigmaz_\sfO ; \hsigmaz_{\sfw }\hsigmaz_{\sfz }\rab
=
\frac{1}{\cZ^2}\bbE\sumtwo{\nu^1\ssim{\lbr\sfO ,\sfw ,\sfz\rbr}\lb\xi^1 ,\frm^1\rb}
{\nu^2\sim  \lb\xi^2 ,\frm^2\rb} \One_{\lbr \sfO\nslra{*}\frg\rbr} .
\ee
In particular, $\partial M/\partial\rho \geq 0$.

{One can readily generalize the latter conclusion
to a system with inhomogeneous flip rates in the following fashion: Let $\rho_\sfe :\bbS_\beta\mapsto \bbR_+;\ \sfe\in \cE^0$ be a collection of
non-negative (and, say, piece-wise smooth)  functions. Let us view the $\rho_\sfe$'s as time-inhomogeneous rates of
arrivals of (ground) flips corresponding to the 
endpoints of $\sfe$. In this way, we may introduce an analog of \eqref{1ptf},
defining $\sfz$-expectation values
\[
M_\sfu \lb \rho (\cdot )\rb  = 
M_\sfu \lb h,  \rho (\cdot ), \lambda \rb;\quad
\sfu\in\frS ,
\]
via the right hand side of \eqref{1ptf} but using the inhomogeneous arrival 
rates $(\rho_e(t))_{e \in \cE^0}$.
}

{Then,
for every $\sfu\in\frS$, the functional $M_\sfu  (\cdot)$ is non-decreasing in $\rho$, that
is
\be
\label{eq:rhoMon}
\forall\sfe\ \rho_\sfe \leq\rho_\sfe^\prime\ \ {\rm t- a.e.}\quad\Longrightarrow\quad
\forall\sfu\ M_\sfu (\rho )\leq M_\sfu (\rho^\prime ) .
\ee
It is worth noting that this may be seen as a special case of Griffith's second inequality \cite{Grif}.}

{Obviously, we may use the random current representation to introduce time-inhomogeneous {versions of} all correlations we have already encountered in this paper.  With that in mind, the following combination of \eqref{eq:rhoMon} with \eqref{eq:Crossmany}
will be useful in the sequel: Let $\frA = \frI_1\cup\dots\cup\frI_n$ be a union of disjoint
ground segments of the form {either for some $i$;  $\frI_l =\bbS_\beta^i$, or} 
$\frI_l = (\sfw_l ,\sfz_l )$, where both {$\sfw_l$ and $\sfz_l$} lie
on some circle $\bbS_\beta^i$ (and are time ordered to avoid notational ambiguities).
{Define ${\frA}^c = \frS\setminus\frA$.}
 Finally define the reduced arrival rates $\rho^\frA$,
\begin{equation}
 \label{eq:ReducedRates}
 \rho^\frA_\sfe (t) =
\begin{cases}
 &\rho ,\ \  \text{if the corresponding flip is either between two points in $\frA$}\\
&\quad \ \  \text{or
between two points in $\frA^c$}\\
& 0 ,\ \ \text{otherwise}.
\end{cases}
\end{equation}
In other words we suppress arrivals of flips between $\frA$ and $\frA^c$.
Let $\sfu\in {\frA^c}$ and let $\sfv_1, \dots ,\sfv_l ,\sfu ,\sfv_{l+1}, \dots ,\sfv_{2n}$
be the time ordering of the set $\lbr \sfw_1, \dots , \sfz_n ,\sfu\rbr$.  Then, exactly as
in \eqref{eq:Crossmany},
\[
 \la \prod_1^l \Hsigmax_{\sfv_q}\hsigmaz_\sfu\prod_{l+1}^{2n} \Hsigmax_{\sfv_q}
\rab (\rho^\frA ) \leq \la\hsigmaz_\sfu\rab (\rho^\frA )
\la \prod_1^{2n}\Hsigmax_{\sfv_q}\rab (\rho^\frA ) \leq M (\rho )
\la \prod_1^{2n}\Hsigmax_{\sfv_q}\rab (\rho^\frA ) ,
\]
where the expectations are understood in terms of the corresponding (generalized to
time-inhomogeneous rates) random current representations, and the second inequality
follows from \eqref{eq:rhoMon}.}

{
In view of how the rates $\rho^\frA$ were defined,
fixing labels at the end-points of $\frI_1 ,\dots ,\frI_n$ completely decouples the two regions $\frA$ and $\frA^c$.  As a result, we obtain the
following inequality:
\be
\label{eq:CondExp}
\bbE_{\rho^{\frA}} \sum_{\nu|_{\check{\frA}^c}\ssim{\sfu}\lb \xi ,\frm \rb }
\prod_1^{2n}\One_{\lbr \nu (\sfv_q ) = r\rbr} \leq M(\rho )
\bbE_{\rho^{\frA}} \sum_{\nu|_{\check{\frA}^c}\sim
\lb \xi ,\frm \rb }
\prod_1^{2n}\One_{\lbr \nu (\sfv_q ) = r\rbr} ,
\ee
where
the expectation above is with respect to $\rho^{\frA}$-arrival rates and 
 the summation is over all  {\textit{reduced}} labels $\nu|_{{{\frA}^c}}:{\frA^c}\mapsto
\lbr r,l\rbr$.}

\section{Differential inequalities}
\label{S:DiffIn}
The following is an adaptation of the ideas of \cite{AB, ABF} to the quantum
case.  It is worth noting that the space-time techniques we develop here yield
simplified proofs even in the classical case.

A fruitful idea of \cite{ABF} is to work with {\em three} replicas in order
to control the above quantities. In our case
these will be three independent replicas $\lb \xi^1 ,\frm^1\rb$,
$\lb \xi^2 ,\frm^2\rb$, $\lb \xi^3 ,\frm^3\rb$
of  Poisson processes of flips and marks and, respectively, three sets
of compatible labels $\nu^1 ,\nu^2, \nu^3$.  We shall always indicate in sub-indices
which replicas we are talking about, e.g. we shall talk about left $l_1$ paths in the
first replica or about unblocked $*_{23}$-paths in the replicas $2$ and $3$.
In the sequel $\bbP$ is the product measure for all three independent replicas and $\bbE$ denotes the corresponding expectation.

Let us go back to the representations \eqref{1ptf} and \eqref{pf},
\be
\label{eq:FullExpression}
 \la \hsigmaz_\sfO\rab = \la \hsigmaz_\sfO\rab\frac{\cZ^2}{\cZ^2} =
\frac1{\cZ^3}\bbE\sum_{\nu^1\ssim{\sfO} (\xi^1 ,\frm^1 )}
\sumtwo{\nu^2\sim (\xi^2 ,\frm^2 )}{\nu^3 \sim (\xi^3 ,\frm^3 )}\1 .
\ee

\begin{figure}[thb] 
\begin{center} 
\psfragscanon 
\psfrag{L}[l]{$\Lambda$}
\psfrag{b}{$\beta$} 
\psfrag{g}{$\frg$}
\psfrag{0}{{\Large$0$}}
\psfrag{t}{{\Large $t$}}
\psfrag{s}{{\Large $s$}}
\psfrag{r}{$r$}
\psfrag{l}{$\ell$}
\psfrag{1}{\case{1}}
\psfrag{2}{\case{2}}
\psfrag{3}{\case{3}}
\psfrag{T1}{$1$-flips}
\psfrag{T2}{$23$-marks}
\psfrag{T3}{Blocked $23$-intervals}
\psfrag{C}{$\Cl^*_{23}(\frg )$}
\includegraphics[width=5in]{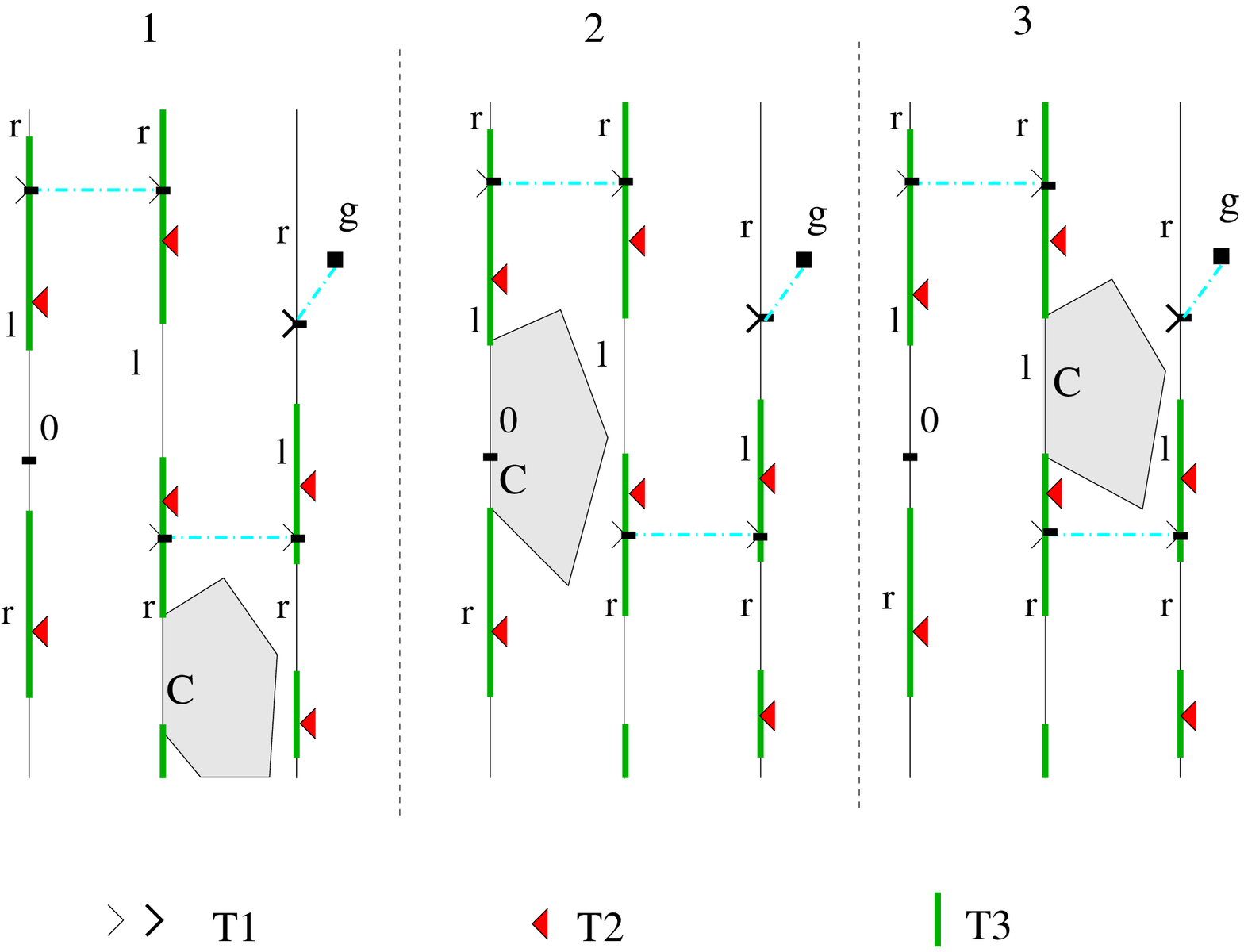}
\end{center} 
\caption{ The ground left $1$-path $\check\Cl^l_1 (\sfO ,\frg )$ contains three intervals. 
$r$ and $\ell$ are $\nu^1$-labels of the first replica. The unblocked 
$23$-cluster $\Cl^*_{23}(\frg )$ is depicted schematically.\newline
 \case{1}:\  $\check\Cl^l_1 (\sfO ,\frg )$ is disjoint from 
 $\Cl^*_{23}(\frg )$.\quad 
\case{2}:\  $0\in \Cl^*_{23}(\frg )$.\newline \case{3}:\ $0\notin \Cl^*_{23}(\frg )$, but 
$\check\Cl^l_1 (\sfO ,\frg )\cap \Cl^*_{23}(\frg )\neq \emptyset$. } 
\end{figure}

Let  $\Cl^*_{23} (\frg )$ be the set
of all
points $\sfv\in\frS$ which are $*_{23}$-connected to $\frg$ and let us denote $\check{\Cl}^l_1 (\sfO ,\frg )$ as the set of ground ($\frS$)
 points on the unique
ground
left path from $\sfO$ to $\frg$. We shall distinguish three cases
which exhaust  all possible contributions to the right hand side of \eqref{eq:FullExpression} and lead to the various terms in \eqref{eq:ThmDifIn1}:
\begin{enumerate}
 \item $\check{\Cl}^l_1 (\sfO ,\frg )\cap\Cl^*_{23} (\frg ) = \emptyset$.
\item $\sfO\in \Cl^*_{23} (\frg )$ .
\item $\sfO\not \in \Cl^*_{23} (\frg )$ but
$\check{\Cl}^l_1 (\sfO ,\frg )\cap\Cl^*_{23} (\frg ) \neq \emptyset$.
\end{enumerate}

Below we consider these cases in turn (see Figure~4 ).  During our exposition of \case{3}, we also derive the pair of inequalities \eqref{eq:ThmDifIn2}.

\case{2} If $\sfO\in \Cl^*_{23} (\frg )$ then there exist $*_{23}$-paths from
$\sfO$ to $\frg$. Hence the notion of the minimal {path} $\cP^* \df \Cl^*_{23}(\sfO ,\frg )$ {from $\sfO$ to $\frg$} is well defined. Applying the Basic Transformation $\Phi_{\cP^*}$ on  $23$-labels, we readily
conclude,
\be
\label{eq:case2}
 \frac1{\cZ^3}\bbE\sum_{\nu^1\ssim{\sfO} (\xi^1 ,\frm^1 )}
\sumtwo{\nu^2\sim (\xi^2 ,\frm^2 )}{\nu^3 \sim (\xi^3 ,\frm^3 )}
\One_{\lbr
\sfO\in \Cl^*_{23} (\frg ) \rbr}
 =
{\frac1{\cZ^3}\bbE\sum_{\nu^1\ssim{\sfO} (\xi^1 ,\frm^1 )}
\sumtwo{\nu^2\ssim{\sfO} (\xi^2 ,\frm^2 )}{\nu^3 \ssim{\sfO} (\xi^3 ,\frm^3 )}\!\!\!
\1 =}
M^3 .
\ee

\case{1}  By construction, $\check\Cl_1^l = \lb \frI_1, \dots ,\frI_{n}\rb$.  All the
intervals in this sequence are ground, and the last interval
$\frI_n = (\sfw ,\sfv \df (i,t ))$ satisfies $t\in \xi^1_{i\frg }$.  
Let $\wt\xi^1$ be the modified
realization of $1$-process of flips with the corresponding arrival removed, but the configuration $(\nu^1, \xi^1, \frm^1)$ otherwise kept intact. Obviously, the relative weight of removing 
this arrival contributes a factor $h \textrm{d} t$, {and one can recover the original 
$\xi^1$ by adding a flip from $\sfv$ to the ghost site $\frg$. 
 Formally, fixing realizations of the second and third replicas and 
fixing compatible values of $\nu^1$ and $\nu^2$, taking expectations {\em only} with
 respect to $(\xi^1 ,\frm^1 )$ and summing
{\em only} with respect to compatible $\nu^1$-labels we obtain }
\begin{multline}
\label{test1}
\mathbb E_1\left[ \sum_{\nu^1 \ssim{\sfO} (\xi^1, \frm^1)} \One_{\{\check\Cl^l_1 (\sfO ,\frg ) \cap
\Cl^*_{23}(\frg) = \varnothing\}} \right] 
\\= \sum_{i \in \mathbb T_N} \int_{0}^{\beta} h \textrm{d} t \mathbb E_1\left[ \sum_{\nu^1 \ssim{\sfO} (\xi^1, \frm^1)}  \One_{\{\check\Cl^l_1 (\sfO ,\frg ) \cap
\Cl^*_{23}(\frg) = \varnothing\}} \One_{\{(i, t) \in \check\Cl^l_1 (\sfO ,\frg )\}}| \xi^1_{i, \frg}(t) =1\right] 
\end{multline}
Now ,
{
\begin{multline}
\label{test2}
 \mathbb E_1\left[ \sum_{\nu^1 \ssim{\sfO} (\xi^1, \frm^1)}  \One_{\{\check\Cl^l_1 (\sfO ,\frg ) \cap
\Cl^*_{23}(\frg) = \varnothing\}} \One_{\{(i, t) \in \check\Cl^l_1 (\sfO ,\frg )\}}| \xi^1_{i, \frg}(t) =1\right] \\
=  \mathbb E_1\left[ \sum_{\nu^1 \ssim{\sfO, \sfv} (\xi^1, \frm^1)} 
\One_{\{ \Cl^l_1 (\sfO ,\sfv  ) \cap
\Cl^*_{23}(\frg) = \varnothing\}}\right] 
\end{multline}
}
with $\sfv= (i, t)$ on the right-hand side.  

Taking into account replicas $2$ and $3$, let us determine the properties of the resulting triple of configurations from the joint integration on the right-hand side of \eqref{test2}.
{ Since  $\Cl^l_1 (\sfO ,\sfv  ) \cap
\Cl^*_{23}(\frg) = \varnothing$}, there exist $*_{12}$-paths from $\sfO$ to $\sfv$
which are disjoint from $\Cl^*_{23} (\frg )$. Let $\cP^*$ be the minimal such path. Consider
the Basic Transformation $\Phi_{\cP^*}$ on $12$-labels. It produces new collection
$\lbr \lb \hat\nu^1 ,\hat\xi^1 ,\hat\frm^1\rb ,\lb \nu^2 ,\hat\xi^2 ,\hat\frm^2\rb\rbr$, which
satisfies the following set of conditions:
\begin{enumerate}
 \item  $\cP^*$ is still the minimal $*_{12}$ path from $\sfO$ to $\sfv$ which avoids
$\Cl^*_{23} (\frg )$. In particular, the transformation is invertible and
$\sf0\nslra{*_{23}}\frg$
\item $\hat\nu^1 \sim \lb \hat\xi^1 ,\hat\frm^1\rb$ and
$\hat\nu^2 \ssim{0, \sfv } \lb \hat\xi^2, \hat\frm^2\rb$. In particular, $\sfO \slra{*_{23}}\sfv$.
\end{enumerate}
Comparing with \eqref{prod2pf_ub} (applied to $2$ and $3$ labels)
and with the first of \eqref{eq:derivatives}, we conclude
\be
\label{eq:case1}
\frac1{\cZ^3}\bbE\sum_{\nu^1\ssim{\sfO} (\xi^1 ,\frm^1 )}
\sumtwo{\nu^2\sim (\xi^2 ,\frm^2 )}{\nu^3 \sim (\xi^3 ,\frm^3 )}
\One_{\lbr
\check{\Cl}^l_1 (\sfO ,\frg )\cap\Cl^*_{23} (\frg ) = \emptyset\rbr}
 \leq \sum_i\int_0^\beta h\dd t\la\hsigmaz_\sfO ;\hsigmaz_{(i,t )}\rab =
h\frac{\partial M}{\partial h} .
\ee

\case{3} This is the most difficult case.  In fact it contains two sub-cases, which we proceed to describe:

The left ground path from $\sfO$ to $\frg$, denoted by  $\check\Cl^l_1 (\sfO ,\frg )$, is a ground path which may be naturally written as
an ordered collection of ground intervals, $\check\Cl^l_1 (\sfO ,\frg )=\cup_1^n \frI_l$:
Each interval $\frI_l\df [ \sfz_l ,\sfw_l]$ is also naturally  {oriented} with respect to the
direction of the path towards $\frg$. Therefore, in the case under consideration we can speak of the first interval  $\frI_{l^*}$ where
$\check\Cl^l_1 (\sfO ,\frg )$ hits $\Cl^*_{23}(\frg )$  and, furthermore about the first
hitting point $\sfu^*\in \frI_{l^*}$.
\smallskip

\noindent
\case{3(a)} {\em Pivotal Marks}:  In this sub-case $ \sfz_{l^*}\nslra{*_{23}}\frg$ or, 
equivalently, $\sfz_{l^*}\neq \sfu^*$. 
Since $\sfu^*$ is in the boundary of $\Cl^*_{23}(\frg )$, there
is a necessarily a $23$-mark at $\sfu^*$.  Also, both the $2$ and $3$ labels are $r$ at $\sfu^*$.
By construction,
(if we understand the interval $\lb\sfz_{l^*},  \sfu^*\rb$ as being topologically {\em open})
\be
\label{eq:Inter1}
\check \Cl^l_1 (\sfO ,\sfu^*)\cap \Cl^*_{23}(\frg ) =\emptyset .
\ee
Hence there exist $*_{12}$-paths from $\sfO$ to $\sfu^*$ which avoid $\Cl^*_{23} (\frg )$.
Let $\cP^*_{12} (\sfO ,\sfu^* )$ be the minimal such path. Let also
$\cP^*_{23} (\sfu^*,\frg )$ be the minimal $*_{23}$-path from $\sfu^*$ to $\frg$.
These paths are disjoint.  Let us make the following {{\em double}} 
transformation on {\em all three}
collections of replicas and compatible labels:
\begin{enumerate}
 \item Remove the $23$-mark at $\sfu^*$.  This yields the weight ${2}\lambda \textrm{d} t$.
\item {Perform the} Basic Transformation $\Phi_{\cP^*_{12} (\sfO ,\sfu^* )}$ on $12$-labels.
\item {Perform the} Basic Transformation $\Phi_{\cP^*_{23} (\sfu^* ,\frg )}$ on $23$-labels.
\end{enumerate}
Since the Basic Transformations are on disjoint paths the latter two operations are well defined, commute and moreover do not change the minimal character of $\cP^*_{12} (\sfO ,\sfu^* )$ and
$\cP^*_{23} (\sfu^*,\frg )$. In other words, they are invertible.  The resulting set
of triples  \newline
$\lbr \lb \hat\nu^1 ,\hat\xi^1 ,\hat\frm^1\rb , \lb \hat\nu^2 ,\hat\xi^2 ,\hat\frm^2\rb,
\lb \hat\nu^3 ,\hat\xi^3 ,\hat\frm^3\rb\rbr$  satisfies the following
 conditions (see Figure~5):
\begin{enumerate}
 \item $\hat \nu^1\ssim{\sfu^*} \lb \hat\xi^1 ,\hat\frm^1\rb$,
$\hat\nu^2\ssim{\sfO}
\lb \hat\xi^2 ,\hat\frm^2\rb$ and $\hat\nu^3\ssim{\sfu^*} \lb \hat\xi^3 ,\hat\frm^3\rb$.
\item $\hat\nu^2 (\sfu^* )=l$.
\item $\sfu^*$ is pivotal for $\lbr\sfO\slra{*_{23}}\frg\rbr$ .
\end{enumerate}
Note that (2) is a consequence of (1) and (3).

\begin{figure}[h] 
\begin{center} 
\psfragscanon 
\psfrag{L}[l]{$\Lambda$}
\psfrag{b}{$\beta$} 
\psfrag{g}{$\frg$}
\psfrag{0}{{\large$0$}}
\psfrag{ls}{$i$}
\psfrag{s}{{\Large $s$}}
\psfrag{r}{$r$}
\psfrag{ldt}{$2\lambda\dd t$}
\psfrag{C}{$\check\Cl^l_1 (\sfO ,\frg )$}
\psfrag{l}{$\ell$}
\psfrag{u}{$\sfu^*$}
\psfrag{v}{$\sfu^*$}
\psfrag{z}{$\sfz^*$}
\psfrag{T1}{$1$-flips}
\psfrag{T2}{$23$-marks}
\psfrag{T3}{(Left) Left ground $1$-path $\check\Cl^l_1 (\sfO ,\frg )$}
\psfrag{T4}{(Right) Minimal $12$-path $\cP^*_{12} (\sfO ,\sfu^* )$}
\psfrag{Cg}{$\Cl^*_{23} (\frg )$}
\psfrag{T}{$\frA^*_{23}(\frg ,\sfu^* )$}
\psfrag{Cv}{$\frA^*_{23}(\sfO ,\sfu^*)$}
\includegraphics[width=5in]{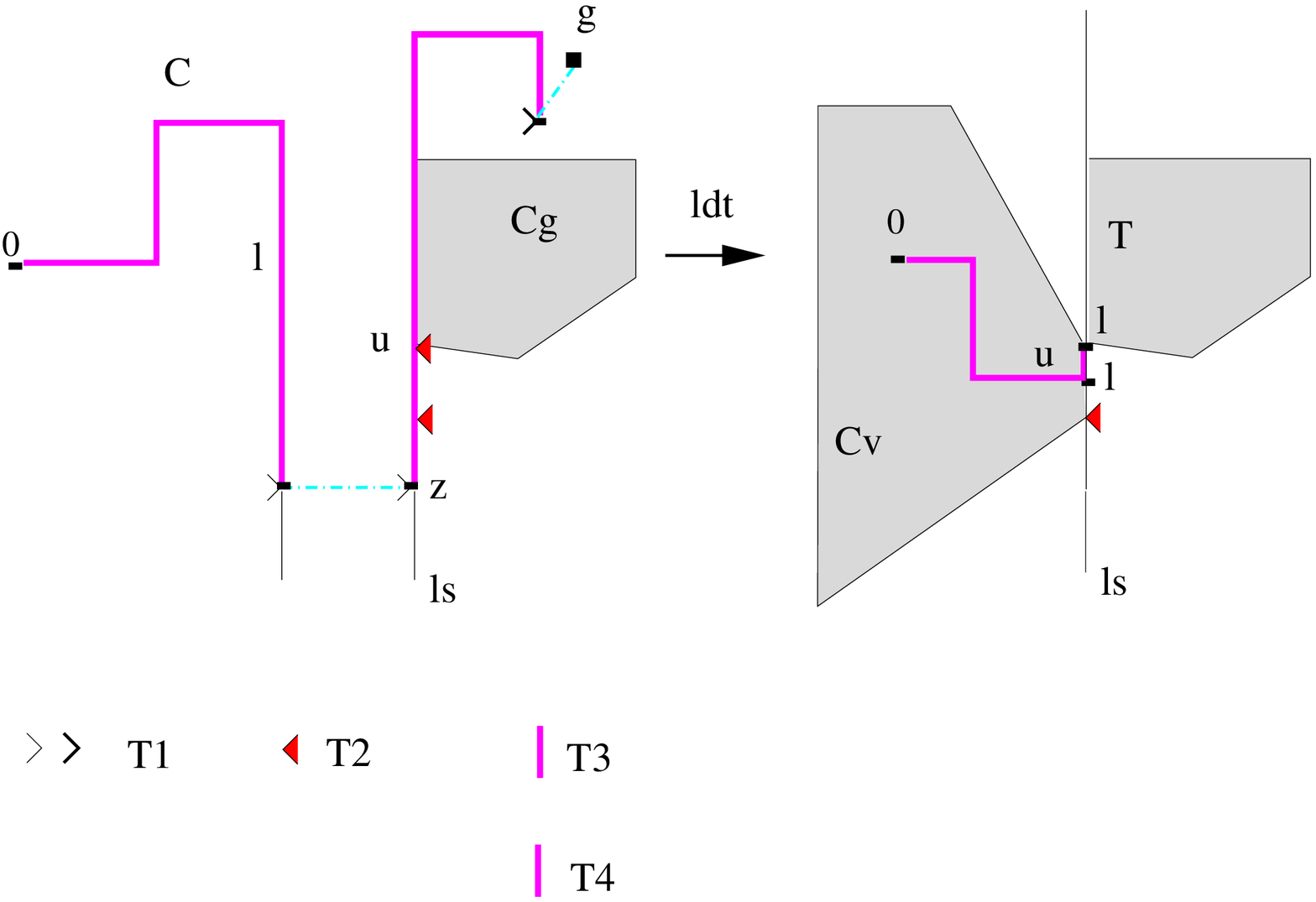}
\end{center} 
\caption{Double transformation in the \case{3 (a)}:  
The $23$-mark at $\sfu^*$ is removed at the cost $2\lambda\dd t$. 
The point $\sfu^*$ is pivotal 
for the $\lbr \sfO\slra{*_{23}}\frg \rbr$-connection in the modified configuration.
$\frA^*_{23} (\sfO ,\sfu^* )$ is  the set of
all the points $\sfu\in\frS$ which can be reached from $\sfO$ via unblocked
$23$-paths {\em avoiding} $\sfu^*$.
$\frA^*_{23} (\frg ,\sfu^* )$ is  the set of
all the points $\sfu\in\frS$ which can be reached from $\frg$ via unblocked
$23$-paths {\em avoiding} $\sfu^*$.
} 
\end{figure}

We claim that,
\be
\label{eq:lTOr}
\begin{split}
&\bbE\!\!\!\!\sumtwo{\nu^2\ssim{\sfO}
\lb \xi^2 ,\frm^2\rb}{\nu^3\ssim{\sfu^*} \lb \xi^3 ,\frm^3\rb}
\!\!\!\!\One_{\lbr \text{$\sfu^*$ is pivotal for $\sfO\slra{*_{23}}\frg$}\rbr}
\One_{\lbr \nu^2 (\sfu^* ) = l\rbr} \\
&\quad \leq M
\bbE\!\!\!\!\sumtwo{\nu^2\ssim{\sfO}
\lb \xi^2 ,\frm^2\rb}{\nu^3\sim  \lb \xi^3 ,\frm^3\rb}
\!\!\!\!\One_{\lbr \text{$\sfu^*$ is pivotal for $\sfO\slra{*_{23}}\frg$}\rbr}
\One_{\lbr \nu^3(\sfu^* ) = r\rbr}.
\end{split}
\ee
Assuming \eqref{eq:lTOr} for the moment, a comparison with \eqref{eq:Trunc2pt_zx} and
with the third of \eqref{eq:derivatives} reveals that the total contribution to $M$ which
comes from the \case{3(a)} is bounded above by
\be
\label{eq:case3a}
M^{2} \frac{1}{\cZ^2}\bbE \sum_i\int_0^\beta {2} \lambda\dd t
\sumtwo{\nu^2\ssim{\sfO}
\lb \xi^2 ,\frm^2\rb}{\nu^3\sim  \lb \xi^3 ,\frm^3\rb}
\!\!\!\!\One_{\lbr \text{$(i, t)$ is pivotal for $\sfO\slra{*_{23}}\frg$}\rbr}
\One_{\lbr \nu^3(\sfu^* ) = r\rbr}  = - 2\lambda
M^{2}\frac{\partial M}{\partial \lambda} .
\ee

To check \eqref{eq:lTOr} let $\frA^*_{23} (\sfO ,\sfu^* )$ be the set of
all the points $\sfu\in\frS$ which can be reached from $\sfO$ via unblocked
$23$-paths {\em avoiding} $\sfu^*$.  Evidently, $\frA^*_{23} (\sfO ,\sfu^* )$
can be written
as a union of intervals $\frA^*_{23} (\sfO ,\sfu^* ) = \cup\cR_j$, which satisfy the following
set of properties:
\begin{enumerate}
 \item Each interval $\cR_j\df (\sfp_j ,\sfq_j )$ (which formally speaking union of successive ground intervals on some $\mathbb S_{\beta}^i$) bears $23$-marks at its endpoints $\sfp_j$ and
$\sfq_j$, except, of course, for the interval which contains
$\sfu^*$ as one of its endpoints --  recall that  the $23$-mark  at $\sfu^*$ was removed.  Moreover, both labels $\nu^2$ and $\nu^3$ equal to $r$ at such 
end-points.
\item Let $\cR_{j^*} = (\sfp_{j^*}, \sfu^*)$ be the remaining interval which contains
$\sfu^*$ as one of its endpoints. Then $\nu^3 (\sfu^*- )\df 
\lim_{\sfz\in\cR_{j^*},  \sfz\to\sfu^*}\nu^3 (\sfz ) = r$.
\item There are no arrivals of $23$-flips between points in $\frA^*_{23} (\sfO ,\sfu^* )$ and
points in $\frS\setminus \frA^*_{23} (\sfO ,\sfu^* )$.
\end{enumerate}

{
The inequality \eqref{eq:lTOr} is then proved as follows: Conditioning on
$ \frA^*_{23} (\sfO ,\sfu^* )$ with realizations of all the
processes and values of both
$2$ and $3$ labels on it, we integrate with respect to marks on
$\frS\setminus \frA^*_{23} (\sfO ,\sfu^* )$, flips on
$\frS\setminus  \frA^*_{23} (\sfO ,\sfu^* )\cup\frg$ and
compatible $2$ and $3$ labels.  The constrained  integration clearly decouples the two configurations on $\frS\setminus  \frA^*_{23} (\sfO ,\sfu^* )\cup\frg$ and so we can integrate the restricted $2$ and $3$
 quantities independently.  We arrive at a situation where \eqref{eq:CondExp} applies (for
the restriction of $\nu^3$). {More precisely, what we use} is actually a limiting case of \eqref{eq:CondExp}, with the $\sfz$ component of spin in the expectation occurring at the 
 point  $\sfu^*$ on the boundary of 
$\frS\setminus  \frA^*_{23} (\sfO ,\sfu^* ){\df \frA^*_{23} (\sfO ,\sfu^* )^c}$.
  Putting things together concludes Step $3(a)$.}
\smallskip

Before proceeding to \case{3(b)}, let us prove the first of \eqref{eq:ThmDifIn2} by techniques similar to those of the previous paragraph.  With the same notation as in Step $3 (a)$, recall
that $\frA^*_{23} (\sfO ,\sfu^* )\cap\Cl^*_{23} (\frg ) = \emptyset$.
Consequently,
 by \eqref{eq:CondExp} (or, more precisely by the limiting case of the latter, 
again applied to the restriction of $\nu^2$ to
$\frA^*_{23} (\sfO ,\sfu^* )^c$ 
at the point $\sfu^*$
on the boundary of $\frA^*_{23} (\sfO ,\sfu^* )^c$),
\[
 \bbE\!\!\!\!\sumtwo{\nu^2\ssim{\sfO}
\lb \xi^2 ,\frm^2\rb}{\nu^3\sim  \lb \xi^3 ,\frm^3\rb}
\!\!\!\!\One_{\lbr \text{$\sfu^*$ is pivotal for $\sfO\slra{*_{23}}\frg$}\rbr}
\One_{\lbr \nu^3(\sfu^* ) = r\rbr} \leq\ M
\bbE\!\!\!\!\sumtwo{\nu^2\ssim{\sfO ,\sfu^*}
\lb \xi^2 ,\frm^2\rb}{\nu^3\sim  \lb \xi^3 ,\frm^3\rb}
\!\!\!\!\One_{\lbr \frA^*_{23} (\sfO ,\sfu^* )\cap\Cl^*_{23} (\frg ) = \emptyset\rbr}
\One_{\lbr \nu^3(\sfu^* ) = r\rbr}.
\]

{
Now 
\[
\bbE\!\!\!\!\sumtwo{\nu^2\ssim{\sfO ,\sfu^*}
\lb \xi^2 ,\frm^2\rb}{\nu^3\sim  \lb \xi^3 ,\frm^3\rb}
\!\!\!\!\One_{\lbr \frA^*_{23} (\sfO ,\sfu^* )\cap\Cl^*_{23} (\frg ) = \emptyset\rbr}
\One_{\lbr \nu^3(\sfu^* ) = r\rbr} \One_{\lbr \sfu^*\nslra{*_{23}}\frg\rbr} \leq \bbE\!\!\!\!\sumtwo{\nu^2\ssim{\sfO ,\sfu^*}
\lb \xi^2 ,\frm^2\rb}{\nu^3\sim  \lb \xi^3 ,\frm^3\rb}
\!\!\!\!
 \One_{\lbr \sfu^*\nslra{*_{23}}\frg\rbr}
 \]
 and the right-hand side is $\mathcal Z^2 \langle \hsigmaz_{\sfO}; \hsigmaz_{\sfu^*}\rangle$ 
(see \eqref{prod2pf_ub}).
On the other hand
\begin{equation}
\label{eq:M2Bound}
\begin{split}
& \bbE\!\!\!\!\sumtwo{\nu^2\ssim{\sfO ,\sfu^*}
\lb \xi^2 ,\frm^2\rb}{\nu^3\sim  \lb \xi^3 ,\frm^3\rb}
\!\!\!\!\One_{\lbr \frA^*_{23} (\sfO ,\sfu^* )\cap\Cl^*_{23} (\frg ) = \emptyset\rbr}
\One_{\lbr \nu^3(\sfu^* ) = r\rbr}\One_{\lbr \sfu^*\slra{*_{23}}\frg\rbr} \\
&\qquad =
\bbE\!\!\!\!\sumtwo{\nu^2\ssim{\sfO}
\lb \xi^2 ,\frm^2\rb}{\nu^3\ssim{\sfu^*}  \lb \xi^3 ,\frm^3\rb}
\!\!\!\!\One_{\lbr \frA^*_{23} (\sfO ,\sfu^* )\cap\Cl^*_{23} (\frg ) = \emptyset\rbr}
\One_{\lbr \nu^2(\sfu^* ) = l \rbr}
\end{split}
\end{equation}
as can be seen by performing our Basic Transformation on the
minimal $*_{23}$ path from $\sfu^*$ to $\frg$ (which would necessarily lie in
$\frA^*_{23} (\sfO ,\sfu^* )^c$).   Since the constraints appearing on the right-hand side imply that  $\sfu^*$ is pivotal, we may use \eqref{eq:lTOr} {and 
bound the right hand-side in 
\eqref{eq:M2Bound} 
by $- M^2\mathcal Z^2 \la \hsigmaz_\sfO ; \Hsigmax_{\sfu^*}\rab$}. 
  The inequality \eqref{eq:ThmDifIn2} then follows easily.}
 
\smallskip

\case{3(b)} {\em Pivotal Flips}: Assume now that $ \sfz_{l^*}\slra{*_{23}}\frg$ or, equivalently,
that $\sfz_{l^*}\in\Cl^*_{23} (\frg )$.
In order to simplify notation set $\sfz^* = \sfz_{l^*}$ and $\sfw^* = \sfw_{l^*-1}$.
 Under the above assumption  $\Cl^*_{23} (\frg )$ is disjoint from
the left path $\Cl^l_1 (\sfO ,\sfw^*)$.
Hence there exist $*_{12}$-paths from $\sfO$ to $\sfw^*$ which avoid
$\Cl^*_{23}(\frg)$.
Let $\cP^*_{12} (\sfO ,\sfw^*)$ be the minimal such path. Let also
$\cP^*_{23} (\sfz^*,\frg )$ be the minimal $*_{23}$-path from $\sfz^*$ to $\frg$.
These paths are disjoint.  Let us make now the following transformation on {\em all three}
replicas and labels:
\begin{enumerate}
 \item Remove  the arrival of $\xi^1$ between $\sfw^*$ and $\sfz^*$, yielding 
the weight $\rho J_{i, j} \textrm d t$.
\item Perform the  Basic Transformation $\Phi_{\cP^*_{12} (\sfO ,\sfw^* )}$ on $12$-labels.
\item Perform the  Basic Transformation $\Phi_{\cP^*_{23} (\sfz^* ,\frg )}$ on $23$-labels.
\end{enumerate}
Again, since the Basic Transformations are on disjoint paths they are well defined and do not change the minimal character of $\cP^*_{12} (\sfO ,\sfw^* )$ and
$\cP^*_{23} (\sfz^*,\frg )$. Thus, they are invertible and the resulting collection
of configurations \newline
$\lbr \lb \hat\nu^1 ,\hat\xi^1 ,\hat\frm^1\rb , \lb \hat\nu^2 ,\hat\xi^2 ,\hat\frm^2\rb,
\lb \hat\nu^3 ,\hat\xi^3 ,\hat\frm^3\rb\rbr$  satisfy the following set of
 conditions (see Figure~6):
\begin{enumerate}
 \item $\hat \nu^1\ssim{\sfz^* }\lb \hat\xi^1 ,\hat\frm^1\rb$,  $\hat\nu^2
\ssim{\lbr \sfO ,\sfw^* ,\sfz^*\rbr}
\lb \hat\xi^2 ,\hat\frm^2\rb$ and $\hat\nu^3\ssim{\sfz^*} \lb \hat\xi^3 ,\hat\frm^3\rb$.
\item $\Cl_{23}^* (\sfO ,\sfw^* )$ and $\Cl_{23}^* (\frg )$ are disjoint.
\end{enumerate}
Therefore, the contribution to $M$ which comes from the \case{3(b)} is bounded by
\be
\label{eq:3bExact}
M\frac1{\cZ^2}\bbE\sum_{i,j}\int_0^\beta  \rho J_{ij}\dd t
\sumtwo{\nu^2\ssim{\lbr\sfO ,(i,t)  ,(j,t)\rbr}\lb \xi^2 ,\frm^2\rb}{\nu^3\ssim{(j,t )}\lb\xi^3 ,\frm^3\rb}
\One_{\lbr \sfO \slra{{*_{23}}} (i,t )\rbr}\One_{\lbr \Cl_{23}^* (\sfO , (i,t) )\cap
\Cl_{23}^* (\frg ) =\emptyset\rbr} ,
\ee

\begin{figure}
\begin{center} 
\psfragscanon 
\psfrag{L}[l]{$\Lambda$}
\psfrag{b}{$\beta$} 
\psfrag{g}{$\frg$}
\psfrag{0}{{\large$0$}}
\psfrag{ls}{$j$}
\psfrag{lj}{$i$}
\psfrag{s}{{\Large $s$}}
\psfrag{w}{$\sfw^*$}
\psfrag{ldt}{$\rho J_{ij}\dd t$}
\psfrag{l}{$\ell$}
\psfrag{u}{$\sfu^*$}
\psfrag{v}{$\sfu^*$}
\psfrag{z}{$\sfz^*$}
\psfrag{C}{$\check\Cl^l_1 (\sfO ,\frg )$}
\psfrag{T1}{$1$-flips}
\psfrag{T2}{$23$-marks}
\psfrag{T3}{Left ground $1$-path $\check\Cl^l_1 (\sfO ,\frg )$}
\psfrag{Cg}{$\Cl^*_{23}(\frg )$}
\psfrag{Cv}{$\Cl^*_{23}(\sfO)$}
\includegraphics[width=5in]{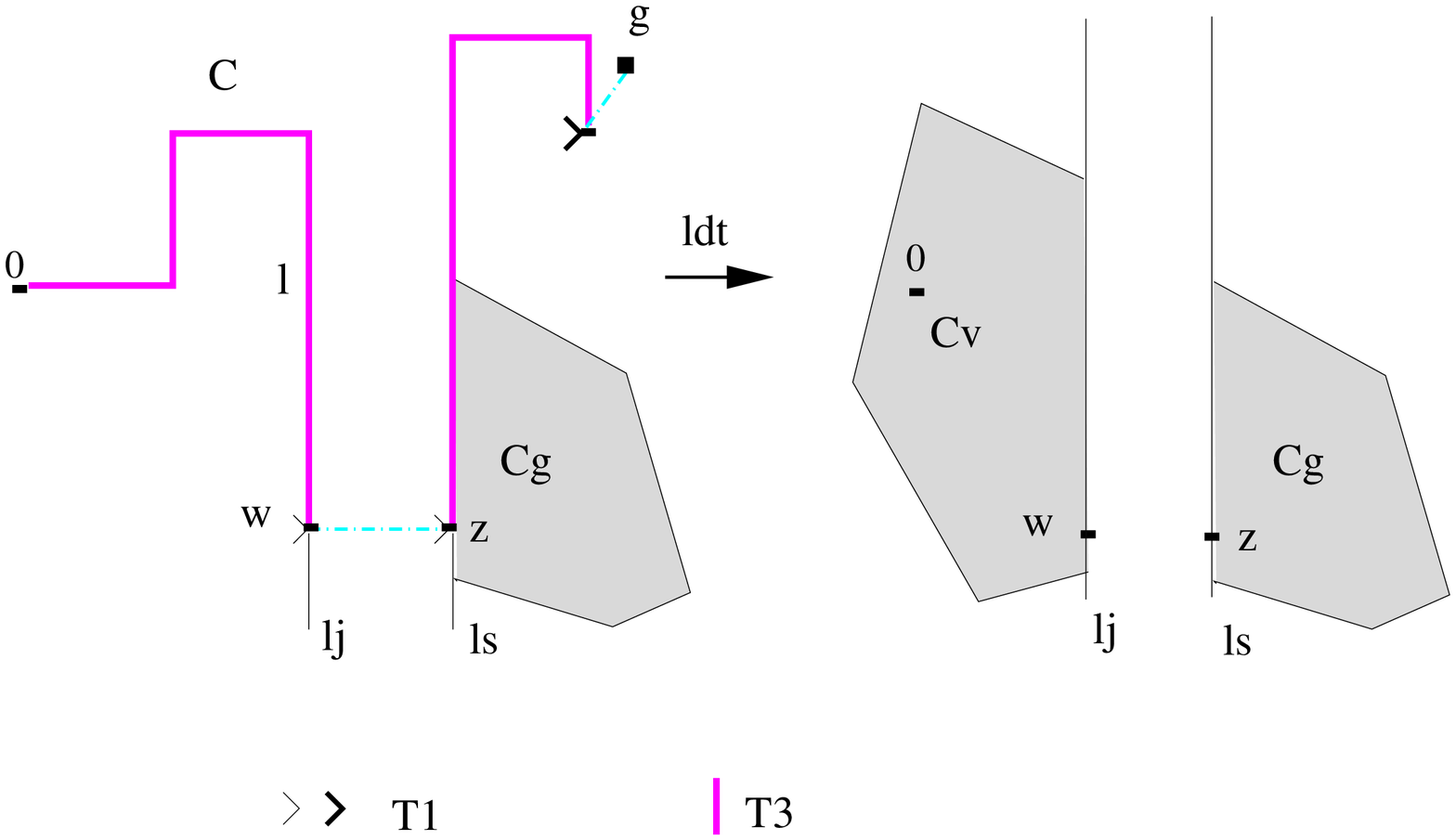}
\end{center} 
\caption{Double transformation in the \case{3 (b)}:  
The $1$-flip between $\sfw^*$  at $\sfz^*$ is removed at the cost $\rho J_{ij}\dd t$. 
In the modified configuration $\sfw^*\in\Cl^*_{23}(\sfO )$ and the clusters 
$\Cl^*_{23}(\sfO )$ and $\Cl^*_{23} (\frg )$ are disjoint.} 
\end{figure}

We claim that the latter expression is bounded above by
\be
\label{eq:3bBound}
 M^2\frac1{\cZ^2}\bbE\sum_{i,j}\int_0^\beta \rho J_{ij}\dd t
\sumtwo{\nu^2\ssim{\lbr \sfO , (i,t) ,(j,t )\rbr}
\lb \xi^2,\frm^2\rb}{\nu^3\sim\lb\xi^3 ,\frm^3\rb}
\One_{\lbr \sfO\slra{{*_{23}}} (i,t )\rbr}\One_{\lbr \Cl_{23}^* (\sfO , (i,t) )\cap
\Cl_{23}^* (\frg ) =\emptyset\rbr} .
\ee
The proof is the same as that of \eqref{eq:lTOr} and is omitted here.  

The expression in
\eqref{eq:3bBound} is exactly $ M^{2} \rho \partial M/\partial\rho$.
Indeed,
just  compare it with \eqref{eq:TripleCor}: If we define $\sfw = (i,t)$ and
$\sfz = (j ,t)$, then
$\sfO\nslra{*_{23}}\frg$ precisely means that either $\sfO\slra{*_{23}}\sfw$,
$\sfz\slra{*_{23}}\frg$  and  $\Cl^*_{23} (\sfO ,\sfw )\cap \Cl^*_{23}(\sfz ,\frg )=\emptyset$
 or, the other way around,
$\sfO\slra{*_{23}}\sfz$,
$\sfw\slra{*_{23}}\frg$  and  $\Cl^*_{23} (\sfO ,\sfz )\cap \Cl^*_{23}(\sfw ,\frg)=\emptyset$.
\smallskip

The second inequality of \eqref{eq:ThmDifIn2} is also an immediate consequence.
 {From a (by now) standard application of the Basic Transformation,
\[
\bbE
\sumtwo{\nu^2\ssim{\lbr \sfO , \sfw , \sfz \rbr}
\lb \xi^2,\frm^2\rb}{\nu^3\sim \lb\xi^3 ,\frm^3\rb}
\One_{\lbr \Cl^*_{23} (\sfO ,\sfw )\cap \Cl^*_{23}(\sfz ,\frg )=\emptyset\rbr }
=
 \bbE
\sumtwo{\nu^2\ssim{\lbr \sfO , \sfw \rbr}
\lb \xi^2,\frm^2\rb}{\nu^3\ssim{\sfz} \lb\xi^3 ,\frm^3\rb}
\One_{\lbr \Cl^*_{23} (\sfO ,\sfw )\cap \Cl^*_{23}(\sfz ,\frg )=\emptyset\rbr }.
\]}
By \eqref{eq:CondExp}
 and in view of the representation \eqref{prod2pf_ub},
\[
 \bbE
\sumtwo{\nu^2\ssim{\lbr \sfO , \sfw \rbr}
\lb \xi^2,\frm^2\rb}{\nu^3\ssim{\sfz} \lb\xi^3 ,\frm^3\rb}
\One_{
\lbr
{ \sfO\nslra{*_{23}}\frg}
\rbr }
\leq
M \bbE
\sumtwo{\nu^2\ssim{\lbr \sfO , \sfw \rbr}
\lb \xi^2,\frm^2\rb}{\nu^3\sim  \lb\xi^3 ,\frm^3\rb}
\One_{\lbr
{ \sfO\nslra{*_{23}}\frg}
\rbr }
 = M \la \hsigmaz_\sfO ;\hsigmaz_\sfw\rab.
\]
The analogous statement holds if the roles of $\sfz$ and $\sfw$ are interchanged.
{The conclusion follows by collecting terms.}

\section{Proof of Theorem A: Exponential  Decay}
\label{S:EXPDEC} 

In the sequel we shall continue to 
use $\bbP$ and, respectively, $\bbE$ for the product probability 
for two independent replicas $\lb \xi^1 ,\frm^1\rb$ and 
$\lb\xi^2 ,\frm^2\rb$. As before $\frn = \frm^1\cup\frm^2$ and 
$\eta = \xi^1\cup\xi^2$.  

The proof is given in three subsections, corresponding to each of the three truncated correlations.  The proof for $\sfz$-correlations is given in some detail and, as the proofs of the second two inequalities only require small modifications of this result, we will be more brief in proving the last two statements.

\subsubsection{Proof of Theorem \ref{Thm:Decay} for $\sfz$-correlations}
Let $i, j \in \mathbb T_N$, $s,t\in\bbS_\beta$  be fixed and let $\sfu= (i, t), \sfv= (j, s)$.
We shall prove the following generalization of the first of \eqref{eq:ThmDecay}  
\begin{lemma}
There 
exist $c_1 =c_1 (h , {\lambda ,\rho } )  >0$  and 
$ c_2 = c_2 (h, {\lambda , \rho }) <\infty$ such that, 
\be 
\label{eq:suvDecay}
\la\hsigmaz_\sfu ;\hsigmaz_\sfv\rab \leq c_2 {\rm e}^{-c_1 \dd (u ,v )} ,
\ee
where $\dd (\sfu ,\sfv )\df |j-i| +|t-s |$.  The above inequality is uniform in $N,\beta, \sfu$ and 
$\sfv$. 
\end{lemma}

\textit{Proof.}
The starting point for our analysis is the formula \eqref{prod2pf_ub} reproduced here:
\begin{equation}
 \label{eq:2ptCore}
 \la\hsigmaz_\sfu ;\hsigmaz_\sfv  \rab  = 
\frac{1}{\cZ^2}\, 
\bbE\lb \sumtwo{\nu^1 \sim (\xi^1 ,\frm^1 )}{\nu^2 \ssim{\sfu,\sfv} (\xi^2 ,\frm^2 )}\One_{\lbr\sfu\slra{* t}\sfv\rbr}
\One_{\lbr \sfu\nslra{*}\frg\rbr}\rb.
\end{equation}
There is a simple 
 reason to include a redundant constraint  $\lbr\sfu\slra{*t}\sfv\rbr$: Given a 
realization of $\xi^1$ and $\xi^2$, the function
\[
 (\frm^1 ,\frm^2)\, \mapsto\, 
\sumtwo{\nu^1\sim (\xi^1 ,\frm^1 )}{\nu^2 \ssim{\sfu,\sfv} (\xi^2 ,\frm^2 )}\One_{\lbr\sfu\slra{* t}\sfv\rbr}
\]
is monotone non-increasing.  Consequently, for any $F (\frm^1 ,\frm^2 )$ non-decreasing, 
the FKG property of  the pair of Poisson processes $\frm^1$ and $\frm^2$ imply:
\begin{multline}
 \label{eq:FKGPoisson}
\bbE \lb F (\frm^1 ,\frm^2 )  \sumtwo{\nu\sim (\xi^1 ,\frm^1 )}{\nu^2\ssim{\sfu,\sfv} (\xi^2 ,\frm^2 )}\One_{\lbr\sfu\slra{*t}\sfv\rbr}\rb 
\\
\leq
\bbE \lb F (\frm^1 ,\frm^2 ) \rb
\bbE \lb  \sum_{\nu^1\sim (\xi^1 ,\frm^1 )}\sum_{\nu^2 \ssim{\sfu,\sfv} (\xi^2 ,\frm^2 )}\One_{\lbr\sfu\slra{*t}\sfv\rbr}\rb.
\end{multline}
\smallskip
For every $\delta >0$ fixed (for convenience we'll assume that $\delta$ divides $\beta$)
{let $\Zdb\df \delta  \bbZ/ \lb (\beta/\delta) \bbZ\rb $ be the rescaled one-dimensional lattice torus
 which is just an equal $\delta$-spacing  embedding of $\beta /\delta $ sites into $\bbS_\beta$ .}
 
 We construct non-decreasing functions 
$F_\delta (\frm^1 ,\frm^2) = F_\delta^{\sfu ,\sfv} (\frm^1 ,\frm^2)$ as follows: 
First of all let us  map $\frS$ onto $\Zdb  \times\bbZ^d$: A point 
$\sfp = (\delta k,j)\in \Zdb  \times\bbZ^d$ corresponds to the interval $[(k-1)\delta ,k\delta )$ of $\bbS^j_\beta$. 
Two points  $\sfp = (\delta k,j)$ and $\sfq = (\delta l, m )$ are said to be connected
if either $j=m$ and {$|k-l|\leq 1\mod ( \beta/\delta )$} or $k=l$ and $(j ,m)\in\cE$.

Consider the following Bernoulli site percolation process $X_\delta$ on 
$\Zdb \times\bbZ^d$, which is generated by the combined process of marks $\frn$
\[
 X_\delta (\sfp  ) =
\begin{cases}
 0,  &\text{if $\frn\lb j\times [(k-1)\delta ,k\delta )\rb >0 $ }\\
\delta , &\text{otherwise}. 
\end{cases}
\]
Clearly,  $\bbP\lb X_\delta = \delta \rb$ tends to one as $\delta$ tends to zero.  For 
$\sfp ,\sfq \in \Zdb \times\bbZ^d$ we can define the minimal passage time
\[
 T_\delta (\sfp, \sfq ) = \min_{\gamma_\delta: \sfp\mapsto \sfq}\sum_{\sfr\in\gamma} X_\delta (\sfr ) .
\]
Then, there exist $c_1, c_2 >0$ such that 
\be 
\label{eq:percBound}
\bbP \lb T_\delta (\sfp, \sfq )  <\frac{\delta}{2} \dd_\delta (\sfp ,\sfq )\rb 
\leq 
c_2{\rm e}^{- c_1\dd_\delta (\sfp ,\sfq )} ,
\ee 
uniformly in $0 \leq \delta \leq \delta_0$ small enough and in $\sfp ,\sfq \in \Zdb \times\bbZ^d$.  Moreover, our choice of $\delta_0$ may be made independent of $\beta$.
Here, $\dd_\delta (\sfp ,\sfq )$ is the minimal possible number of points in 
connected paths $\gamma_\delta: \sfp\mapsto \sfq$.   

Note that 
if $\sfp_\sfu$ and $\sfp_\sfv$ label $\delta$-intervals containing $\sfu$ and $\sfv$, then
\[
 \dd_\delta (\sfp_\sfu ,\sfp_\sfv ) \geq c_3 \dd (\sfu ,\sfv )
\]
uniformly in $\delta$ small and, say, $\dd (\sfu ,\sfv )\geq 1$.  Suppose that for such 
$\delta, \sfp_\sfu$ and $\sfp_\sfv$ , we also assume $\delta> 0$ is chosen to satisfy \eqref{eq:percBound}.  If we define
\[
\cD^c_\delta = \lb  \cD_\delta^{\sfu ,\sfv}\rb^c   = \lbr 
T_\delta (\sfp_\sfu, \sfp_\sfv )  <\frac{\delta}{2} \dd_\delta (\sfp_\sfu ,\sfp_\sfv )\rbr ,
\]
then
since 
 $F_\delta \df\One_{\cD^c_\delta} $ is non-decreasing,  the FKG inequality 
\eqref{eq:FKGPoisson} along with \eqref{eq:percBound} imply  that for all $\delta$ small there
exist $c_1= c_1 (\delta ), c_2 >0$, such that 
\begin{equation}
 \label{eq:NoD}
\bbE\lb \One_{\cD^c_\delta  } \sumtwo{\nu^1\sim (\xi^1 ,\frm^1 )}{\nu^2\ssim{\sfu,\sfv} (\xip ,\frmp )}\One_{\lbr\sfu\slra{*t}\sfv\rbr}\rb \leq c_2
{\rm e}^{-c_1  {\sf d} (\sfu ,\sfv )} 
\bbE\lb  \sumtwo{\nu^1\sim (\xi^1 ,\frm^1 )}{\nu^2 \ssim{\sfu,\sfv} (\xi^2 ,\frm^2 )}\One_{\lbr\sfu\slra{*t}\sfv\rbr}\rb .
\end{equation}

In view of \eqref{eq:NoD} it suffices
to check that, perhaps by adjusting further $c_1, c_2 >0$, 
\begin{equation}
 \label{eq:OnD}
\bbE\lb \One_{\cD_\delta} \!\!\!\sumtwo{\nu^1\sim (\xi^1 ,\frm^1 )}{\nu^2\ssim{\sfu,\sfv} (\xi^2 ,\frm^2 )}
\One_{\lbr\sfu\slra{*t}\sfv\rbr}
\One_{\lbr \sfu\nslra{*}\frg\rbr}\rb \leq
c_2{\rm e}^{-c_1 {\sf d} (\sfu ,\sfv )}
\bbE\lb
\One_{\cD_\delta} \!\!\!\sumtwo{\nu^1\sim (\xi^1 ,\frm^1 )}{\nu^2\ssim{\sfu,\sfv} (\xip ,\frmp )}
\One_{\lbr\sfu\slra{*t}\sfv\rbr}\rb .
\end{equation}
Consider now the set $\frA^{*\frt} (\sfu ,\sfv )$ of all the points $\sfz\in\frS$ which are 
$*\frt$-connected to both $\sfu$ and $\sfv$ (see above Figure 2 for the definition). The set  $\frA^{*\frt} (\sfu ,\sfv )$ is non-empty 
on the event $\lbr \sfu\slra{*t}\sfv\rbr$, and it is represented as a union of intervals 
$\frA^{*\frt} (\sfu ,\sfv ) = \cup_l \frR_l$.  Each interval $\frR_l\subset\bbS_\beta^l$ is either
empty (this case is included for notational convenience), or it is a full circle, 
or $\frR_l = (\sfz_l ,\sfw_l )\subset \bbS_\beta^{i_l}$ with combined $\frn$ marks 
placed at both end-points (it could happen that $\sfz_l = \sfw_l$, of course).  Note that these endpoints must also have $\nu^1, \nu^2 = r$.

 Let us say that $\sfp = (\delta k, l)\in\frG_\delta (\frR_l )$ if
\[
 \frR_l\neq\emptyset ,\ \sfp\in \frR_l\quad\text{and}\quad X_\delta (\sfp )=\delta .
\]
Note that $\sfp\in{\frG_\delta}(\frR_l)$ implies in particular that $[(k-1)\delta , k\delta) \times l \subseteq\frR_l$.

The crucial property
is that on the event $\cD_\delta^{\sfu ,\sfv}$
the following happens:
 The number of 
all $\delta$-intervals associated with points $\sfp\in\cup_l \frG_\delta (\frR_l )$ 
 is bounded below as
\be  
\label{eq:AstBound}
\sum_l\sum_{\sfp\in \frG_\delta (\frR_l) }\1  \geq \frac1{\delta} T_\delta (\sfp_\sfu  ,\sfp_\sfv ) >  
c_3 \frac{1}{2}\dd (\sfu ,\sfv ) .
\ee

Let us condition on realizations of $\frA^{*\frt} (\sfu ,\sfv )$ which are compatible 
with $\lbr \sfu\slra{*t}\sfv\rbr$ and $\cD_\delta^{\sfu ,\sfv}$.   As before, such a conditioning
rules out simultaneous flips between points in $\frA^{*\frt} (\sfu ,\sfv )$ and 
$\frS\setminus\frA^{*\frt} (\sfu ,\sfv )$.  Therefore, the corresponding 
conditional integration and summation over
compatible flips, marks and labels inside and outside 
$\frA^{*\frt} (\sfu ,\sfv )$ decouples over the two regions.  

In other words, to establish 
\eqref{eq:OnD} it is enough to prove the following statement: Let $\frA = \cup\frR_l$ 
be a collection of disjoint intervals, such that 
$\sfu$ and $\sfv$ are interiour points of $
\frA$. 
Further, suppose that $
\frA$ contains at least $c_3 \frac{1}{2}\dd (\sfu ,\sfv )$ disjoint sub-intervals each with length at least $\delta$ {and let us say that 
$\cD_\delta^{\sfu ,\sfv } (
\frA )$ occurs for the realization of the combined process of 
marks $\frn$ whenever \eqref{eq:AstBound} holds.}

 Let $ \rho^{
\frA}$ denote the reduced time-inhomogeneous 
rates of arrivals of flips (associated to edges on the torus) as in \eqref{eq:ReducedRates}, 
\begin{equation}
 \label{eq:ReducedRatesg}
 {\rho^{
\frA}}_\sfe (t) =
\begin{cases}
 &\rho ,\ \  \text{if the corresponding flip is either between two points in $
\frA
$}\\
&\quad \ \  \text{or
between two points in $
\frS\setminus 
\frA
$}\\
& 0 ,\ \ \text{otherwise}.
\end{cases}
\end{equation}
Then, 
\begin{equation}
\begin{split}
 \label{eq:OnDReduced}
&\bbE_{\rho^{
 \frA}}
\!\!\!
\sumtwo{\check\nu^1\sim (\xi^1 ,\frm^1 )}{\check \nu^2\ssim{\sfu,\sfv} (\xi^2 ,\frm^2 )}
\One_{\lbr\Cl^{*\frt}(\sfu ,\sfv )= 
\frA \rbr}
\One_{\lbr \sfu\nslra{*}\frg\rbr} {\One_{\lbr\cD_\delta^{\sfu ,\sfv } (%
\frA )\rbr}}
\\
&\qquad  \leq
c_2{\rm e}^{-c_1 {\sf d} (\sfu ,\sfv )}
\bbE_{\rho^{
\frA}}
\!\!\! 
\sumtwo{\check \nu^1\sim (\xi^1 ,\frm^1 )}{\check \nu^2\ssim{\sfu,\sfv} (\xip ,\frmp )}
\One_{\lbr \Cl^{*\frt}(\sfu ,\sfv )= 
\frA\rbr} 
 {\One_{\lbr\cD_\delta^{\sfu ,\sfv } (
\frA )\rbr}},
\end{split}
\end{equation}
where $\check\nu^1 ,\check\nu^2$ are restrictions of the labels to $
\frA = \cup\frR_l$ 
which are
compatible with the marks, and in particular with $r$-boundary conditions, at the end-points of
$\frR_l$-s.  Above $\Cl^{*\frt}(\sfu ,\sfv )$ is the set of points which are $*\frt$-connected to
$\sfu$ and $\sfv$.

The inequality \eqref{eq:OnDReduced} is established by the following embedding procedure:
Let $\lbr \lb \check\nu^1 ,\xi^1 ,\frm^1 \rb , \lb \check\nu^2 ,\xi^2 ,\frm^2 \rb\rbr$ be a pair of configurations which contribute to the left hand side of \eqref{eq:OnDReduced}.  All such 
configurations  have no arrivals of $\frg$-induced flips on $
 \frA$.  At this 
stage it is 
convenient to introduce the following separate notation for processes of flips: let $\check\xi_\sfe^k;
 k=1,2, $
to denote arrivals for $\sfe = (i,j)\in\cE^0$ and $\xi^{\frg ,k}_\sfe; k=1,2, $ to denote arrivals for 
$\sfe =(i,\frg )\cE^\frg$.  A similar notation $\check\eta = \check\xi^1\cup\check\xi^2$ and 
$\eta^\frg = \xi^{\frg ,1} \cup \xi^{\frg ,2}$ is introduced for combined processes of flips.
Then, on the event $\eta^\frg \lb \frA \rb = \emptyset$, the compatibility conditions
on the left hand side of \eqref{eq:OnDReduced} read as
$\check\nu^1\sim \lb \check\xi^1 ,\frm^1\rb$ and, accordingly, 
$\check\nu^2\ssim{\sfu ,\sfv} \lb \check\xi^2 ,\frm^2\rb$, and the expression on the 
left hand side of  \eqref{eq:OnDReduced} equals to
\be 
\label{eq:LeftHside}
{\rm e}^{-2h\sum_l |\frR_l |} \bbE_{\rho^\frA}
\!\!\!
\sumtwo{\check\nu^1\sim (\check\xi^1 ,\frm^1 )}{\check \nu^2\ssim{\sfu,\sfv} 
(\check\xi^2 ,\frm^2 )}
\One_{\lbr\Cl^{*\sft}(\sfu ,\sfv )= \frA\rbr}
{\One_{\lbr\cD_\delta^{\sfu ,\sfv } (\frA )\rbr}}.
\ee

Fix now a realization 
of $(\check\xi^1 ,\frm^1), (\check\xi^2 ,\frm^2 )$ and compatible labels 
$\check\nu^1\sim (\check\xi^1 ,\frm^1 )$ and $\check \nu^2\ssim{\sfu,\sfv} 
(\check\xi^2 ,\frm^2 )$.   Consider the following event
\[
 \frE (\frA ) = \cap_{l}\cap_{\sfp\in\frG_\delta (\frR_l )}\lbr \xi^{\frg ,i}(\frI_\sfp )\ \text{is even 
for $i=1,2$}\rbr\cap\lbr\eta^\frg (\frA\setminus \frA_\delta ) = 0\rbr, 
\]
where, for $\sfp = (k\delta, l )\in \delta \Zdb \times \bbZ^d$, 
 we set 
\[
\frI_\sfp =  [(k-1)\delta ,k\delta )\times l\quad \text{ and}\quad   \frA_\delta = 
\cup_l\cup_{\sfp\in\frG_\delta (\frR_l )}\frI_\sfp .
\]
  Evidently, 
\be  
\label{eq:Intervals}
\bbP\lb \frE (\frA )\rb = 
{\rm e}^{-2h\sum_l |\frR_l |}\prod_{l}\prod_{\sfp\in\frG_\delta (\frR_l )}
\lb\cosh (\delta h )\rb^2 \geq 
{\rm e}^{-2h\sum_l |\frR_l |}
\lb\cosh (\delta h )\rb^{c_3\dd (\sfu ,\sfv )}
\ee
where the second inequality follows from \eqref{eq:AstBound}.  Each  
$\frE (\frA )$-realization of 
$\lb \xi^{\frg , 1}, \xi^{\frg ,2 }\rb $ gives rise to compatible labels 
$ \check\nu^1 [ \xi^{\frg , 1} ]\sim \lb \check\xi^1 , \xi^{\frg , 1}, \frm^1\rb$ and 
$ \check\nu^2 [ \xi^{\frg , 2} ]\ssim{\sfu ,\sfv} \lb \check\xi^2 , \xi^{\frg , 2}, \frm^2\rb$ which 
are unambiguously constructed from the original
$\check\nu^1\sim (\check\xi^1 ,\frm^1 )$ and $\check \nu^2\ssim{\sfu,\sfv} 
(\check\xi^2 ,\frm^2 )$ 
 by the appropriate
even number of flips on each of the intervals $\frI_\sfp\subseteq\frA_\delta$ 
(see Figure~7). 

\begin{figure}[h] 
\begin{center} 
\psfragscanon 
\psfrag{L}[l]{$\Lambda$}
\psfrag{b}{$\beta$} 
\psfrag{g}{$\frg$}
\psfrag{0}{{\large$0$}}
\psfrag{ls}{$i$}
\psfrag{s}{{\Large $s$}}
\psfrag{r}{$r$}
\psfrag{ldt}{$2\lambda\dd t$}
\psfrag{C}{$\check\Cl^l_1 (\sfO ,\frg )$}
\psfrag{l}{$\ell$}
\psfrag{u}{$\sfu$}
\psfrag{v}{$\sfv$}
\psfrag{z}{$\sfz^*$}
\psfrag{R1}{$\frR_1$}
\psfrag{R2}{$\frR_2$}
\psfrag{R3}{$\frR_3$}
\psfrag{X}{Intervals with $X_\delta = 0$}
\psfrag{A}{Ground flips and marks $\lb\check\xi^1 ,\frm^1\rb$ }
\psfrag{B}{Ground flips and marks $\lb\check\xi^2 ,\frm^2\rb$ }
\includegraphics[width=4.5in]{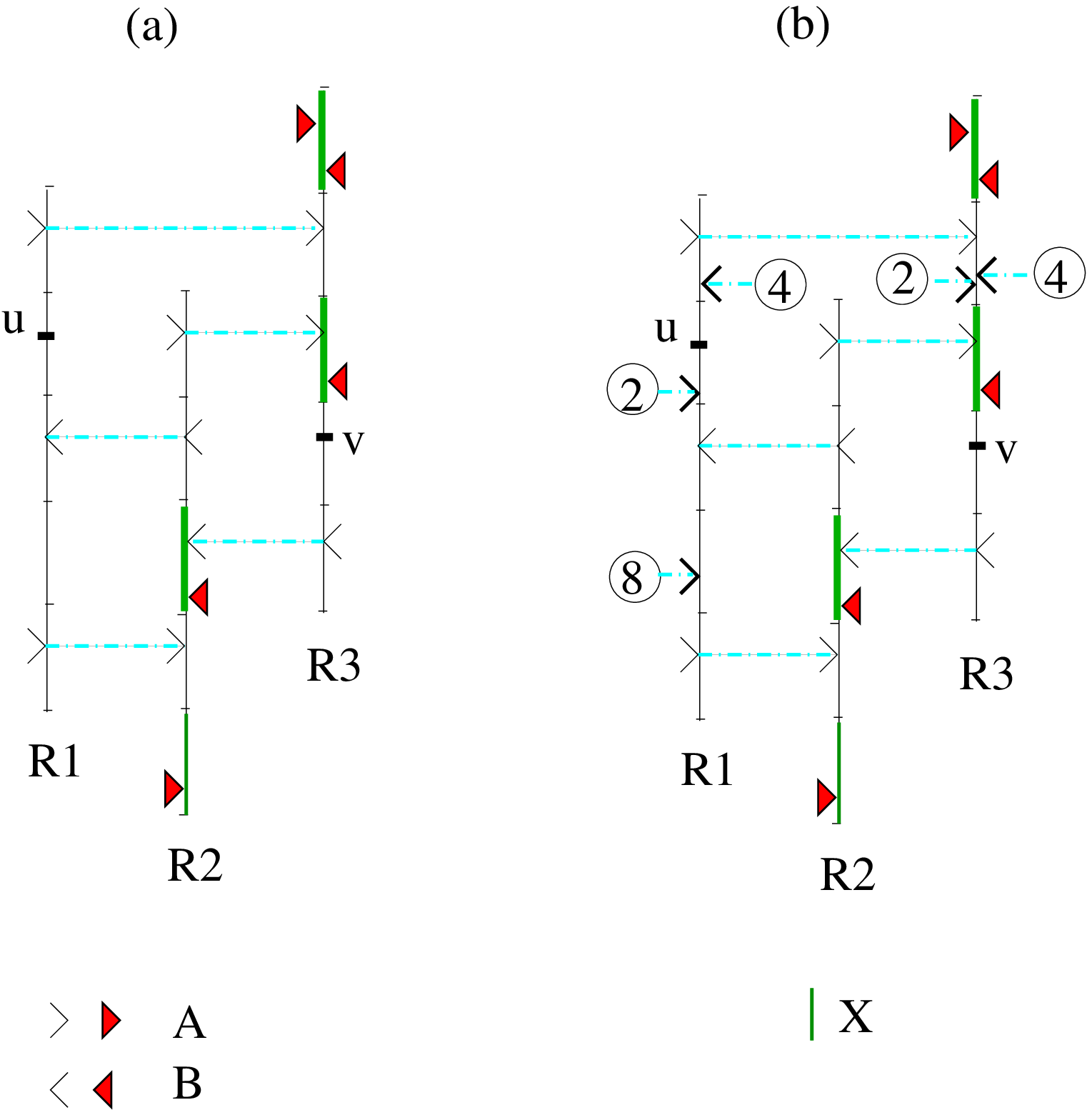}
\end{center} 
\caption{Compatible labels which are constructed from $\check\nu^1$, $\check\nu^2$ and
even number of arrivals of $\xi^{\frg ,1}$ and $\xi^{\frg ,2}$ on intervals 
 from $\frG_\delta$:
\newline 
 (a) Original configurations $\lb \check\nu^1 ,\check\xi^1,
\frm^1\rb$ and $\lb \check\nu^2 ,\check\xi^2,
\frm^2\rb$. 
\newline
  (b) Example of admissible (in the sense of event $\frE$) even number of arrivals of 
$\lb \xi^{\frg ,1} ,\xi^{\frg ,2 }\rb$: the circled numbers indicate total number of 
arrivals on the corresponding intervals.}
\end{figure}

As a result, the expectation on the {right-hand} 
side of \eqref{eq:OnDReduced} is bounded
below by
\[
\lb\cosh (\delta h )\rb^{c_3\dd (\sfu ,\sfv )} 
{\rm e}^{-2h\sum_l |\frR_l |} \bbE_{\rho^\frA}
\!\!\!
\sumtwo{\check\nu^1\sim (\check\xi^1 ,\frm^1 )}{\check \nu^2\ssim{\sfu,\sfv} 
(\check\xi^2 ,\frm^2 )}
\One_{\lbr\Cl^{*\sft}(\sfu ,\sfv )= \frA\rbr}
{\One_{\lbr\cD_\delta^{\sfu ,\sfv } (\frA )\rbr}} , 
\]
and \eqref{eq:OnDReduced} follows.\qed

\subsubsection{Proof of Theorem A for $\sfx \sfz$-correlations}
Recall the expression \eqref{eq:Trunc2pt_zx}, 
\begin{equation}
\label{eq:Trunc2pt_zxP}
\la\hsigmaz_\sfu ; \Hsigmax_\sfx\rab = 
 - \frac1{\cZ^2}\ 
\bbE\!\!\!\!
 \sumtwo{\nu^1\ssim{\sfu} (\xi^1 ,\frm^1 )}{\nup\sim (\xip ,\frmp )}
\One_{\lbr \nu^2 (\sfv ) = r\rbr}
\One_{\lbr \text{ $\sfv$ is pivotal for $\sfu\slra{*}\frg$ }\rbr}.
\end{equation}

Observe that if  $\sfv$ is pivotal for $\sfu \slra{*}\frg$ 
then $\frA^* (\sfu ,\sfv )$  (recall that the latter notation stand for the set of points 
which are $*$-connected to $\sfu$ by paths {\em avoiding} $\sfv$) 
does not contain $\frg$, which means that there are no arrivals of $\eta^\frg$ on 
$\frA^* (\sfu ,\sfv )$. At this point we may  proceed exactly as in the proof of 
Theorem A for $\sfz $-correlations.
\qed

\subsubsection{Proof of Theorem \ref{Thm:Decay} for $\sfx$-correlations}
Recall the expression \eqref{eq:Trunc2pt_x}
\begin{equation}
\la\Hsigmax_\sfu ; \Hsigmax_\sfv\rab  = 
\frac1{\cZ^2}\ 
\bbE 
\!\!\!\!
\sumtwo{\nu^1\sim (\xi^1 ,\frm^1 )}{\nup\sim (\xip ,\frmp )}
\One_{\lbr (\nu^1 , \nu^2 )\in [(r,l), (r,l )]\rbr}\One_{\{\sfv \text{ is loop pivotal for $\sfu$}\}} .
\end{equation}
Observe that under the constraints on the right-hand side, if $\sfv$ is loop-pivotal for $\sfu$ 
 then the set $\frA^* (\sfu ,\sfv )\backslash \{\sfu, \sfv\}$ contains at least two disjoint components. 
 Hence at least
one of these components should be disjoint from $\frg$. Again, at this point 
we may  proceed exactly as in the proof of 
Theorem A for $\sfz $-correlations.
\qed

\subsubsection{Implications for the ground state $\beta =\infty$} As was proved above, exponential decay 
of truncated two-point functions is uniform in $\beta <\infty$.  Consequently,  for 
every $N <\infty$, the limit 
\[
 M_{\infty , N} (h, \rho ,\lambda ) \df \lim_{\beta\to\infty } M_{\beta ,N} (h, \rho ,\lambda )
\]
also satisfies \eqref{eq:ThmDifIn1} and  \eqref{eq:ThmDifIn2}. On the other hand, by 
an obvious time scaling, $M_{\infty , N} (\alpha h, \alpha \rho ,\alpha \lambda ) = 
M_{\infty , N} (h, \rho ,\lambda )$ for every $\alpha >0$.  Hence, 
\[
{\rho} \frac{\partial M_{\infty ,N}}{\partial \rho} = - {\lambda} \frac{\partial M_{\infty ,N}}{\partial \lambda } - 
 {h} \frac{\partial M_{\infty ,N}}{\partial h}\leq 
-  {\lambda} \frac{\partial M_{\infty ,N}}{\partial \lambda }
\]
Therefore, \eqref{eq:ThmDifIn1} implies that 
\be  
\label{eq:Dif1Inf}
M_{\infty ,N}  
 \leq h\frac{\partial M_{\infty ,N}}{\partial h} + M_{\infty ,N}^3 -
{3} M_{\infty ,N}^{2} \lambda  \frac{\partial M_{\infty ,N}}{\partial\lambda} .
\ee
Together with the first of \eqref{eq:ThmDifIn2} (for $M_{\infty ,N}$) the inequality
\eqref{eq:Dif1Inf} sets up the stage for an analysis of sharpness of
 of the $\hsigmaz$ 
 phase transition 
literally along the lines of \cite{AB, ABF}.

\noindent
{
{\bf Acknowledgement.} 
Our proof of exponential decay is based on an argument which was developed 
in the classical setting together  with Roberto Fernandez and Yvan Velenik (see 
\cite{ILN}). We are grateful to Anna Levit for useful remarks and 
a very careful reading  of
the first draft of this paper.}

\end{document}